\documentclass{article}
\usepackage{graphicx} 

\usepackage{appendix}
\usepackage{amssymb}
\usepackage{soul}
\usepackage{xcolor}
\usepackage{amsmath} 
\usepackage{subfigure}
\usepackage{graphicx}
\usepackage[font=small]{caption}
\newcommand{\Var}{\textrm{Var}}
\newcommand{\calM}{\mathcal{M}}
\newcommand{\R}{\mathbb{R}}

\DeclareMathSymbol{\shortminus}{\mathbin}{AMSa}{"39}

\title{Tests of thermal macroeconomic theory on simulated {micro-}economies}

\author{Yihang Luo\footnote{Warwick Business School, yihangluo99@gmail.com, ORCID 0009-0005-8818-2420}, R.S.MacKay\footnote{Mathematics Institute, R.S.MacKay$@$warwick.ac.uk, ORCID:~0000-0003-4771-3692}, 
Nick Chater\footnote{Behavioural Science Group, Warwick Business School, Nick.Chater$@$wbs.ac.uk, ORCID:~0000-0002-9745-0686} \\
 \\
University of Warwick, Coventry CV4 7AL, U.K.
}

\date{\today}

\begin{document}

\maketitle

\begin{abstract}
In this paper, we test predictions of a new theory of macroeconomics, called ``thermal macroeconomics.'' The theory aims to apply the mathematical structure of classical thermodynamics, including analogues of temperature and entropy, to predict aspects of the aggregate behaviour of populations of economic agents without analyzing their detailed interactions. We test the theory by comparing its predictions with the behaviour of a variety of simulated micro-economies in which goods and money can be exchanged between agents, confirming the predictions of the theory. The paper serves also to illustrate and make more tangible the predictions of thermal macroeconomics.  
\end{abstract}

\noindent {\em keywords}:~thermal macroeconomics, agent-based simulation, Carnot cycle, flexibility matrix, Onsager matrix, trade

\section{Introduction}

The paper \cite{CM} introduces a thermodynamic approach to macroeconomics, called ``thermal macroeconomics''.\footnote{We initially called our theory ``thermoeconomics'' but found that this term had already been used in a somewhat different sense.  We note that our account is not a theory of the economics of heat, but an application of the formal structure of thermodynamics to economic phenomena.}  Thermal macroeconomics is ``thermal'' because the mathematical structure has close parallels with classical thermodynamics (the theory of heat and work, temperature, pressure, volume, chemical potentials and so on).
It is ``macro'' in that it focuses on relationships between macroscopic quantities such as quantities and prices of goods, and makes no {specific} assumptions about ``microfoundations'' in terms of the preferences or decision-making strategies of individual agents, { beyond a list of reasonable axioms} {defined at the macro-level}. 
This parallels thermodynamics, which makes no model of microfoundations of physical systems {apart from some plausible axioms} {about the behaviour of macroscopic quantities}.
In{deed, in} physics, the macro-description of classical thermodynamics pre-dated the molecular theory of matter. 

The potential appeal of thermal macroeconomics is that it might, similarly, provide a powerful macro-level theory of the economy, without having to make specific micro-foundational assumptions about economic agents and their interactions. This is particularly attractive because such details are likely to lead to extreme complexity if we move away from idealized rational decision-makers to more realistic cognitive  agents engaged in real-world transactions. 

To state it in a little more detail, the hope is that thermal macroeconomics provides a useful description of macro-level behaviour, with relations that are independent of detailed microfoundational assumptions.  That is, as in the case of thermodynamics in physics, while microfoundational assumptions will of course {affect} macroscopic behaviour, the macroscopic quantities (here prices, quantities of goods, etc) can be measured and related according to the theory, without needing to know {anything about} the underlying microfoundations. In particular, the theory predicts that each sufficiently large economic system possesses a function of statistical state called entropy such that when a set of economies is put in any form of contact, the total entropy can not decrease.  The entropy function depends on the micro-structure of the economy, but the law of non-decrease does not.

Thermal macroeconomics has so far been developed only for so-called ``exchange economies,'' in which infinitely divisible durable goods of finitely many types can be exchanged but there is no production or consumption. So we restrict attention to such systems {here}.  One of the goods is called money and is required to be ``desirable,'' at least at the macroscopic level:~that is, at the macro-level more money is always preferred to less. 

Here, we test the theory by simulations of economies of agents interacting under various forms of stochastic micro-dynamics. The aim is to illustrate {and test} the implications of thermal macroeconomic theory for concrete examples of simulated economies.

\section{Simulated economies}

Although there are now sophisticated agent-based models of economies \cite{AF}, we focus on simple cases of exchange economies, where we can test the detailed predictions of thermal macroeconomics.

The starting point for building our simulated micro-models of exchange economies, against which thermal macroeconomics can be tested, was introduced in \cite{CM}.  In the simplest case, the model for an exchange economy consists of a number $N$ of agents, and positive real {quantities} $M$ of money and $G$ of one other type of good, but extension to more types of good is straightforward.  Following \cite{CM}, we call the unit of money the ``aurum'' and use the ancient symbol $\odot$ for this.  

Each agent has a non-negative ``utility'' function $u(g,m)$ for the {quantities} $g$ of goods and $m$ of money that it owns.  The basic utility function that is our starting point in these simulations has {so-called} Cobb-Douglas form 
\begin{equation}
u(g,m) = g^{\alpha-1}m^{\eta-1}
\end{equation}
for some real exponents $\alpha, \eta > 0$ (there are various other {possible} conventions, e.g.~without the minus ones and/or taking the logarithm, but it is convenient to fix on this one).   
For example, one can take $\alpha=2$, $\eta = 3$, {but non-integer values are allowed too}. 
As a refinement, we allow different agents $i$ to have different exponents $\alpha_i, \eta_i$.  We will also allow different forms of utility function, including {dependence between the {quantities} of goods and money,} and dependence on the holdings of other agents. {Such utility functions capture what are known in behavioural science as \emph{social} preferences, rather than agents caring only about their own holdings of goods and money.}

In contrast to micro-foundations in standard economic theory, our agents are not strict utility-maximisers. Instead, their utility functions bias the outcome of encounters with other agents {probabilistically}---each agent is biased towards a redistribution in proportion to that agent's utility after the redistribution. If desired, utility-maximisation can be approached by raising the utility functions to a large power. 

We take the agents to encounter each other independently at rates given by a symmetric matrix $k_{ij}$ with non-negative entries, representing the probability per unit time at which agent $i$ encounters agent $j$.  A method to simulate a sequence of encounters is described in Appendix~\ref{app:A}.  We rule out self-encounters by taking $k_{ii}=0$.  The simplest example is fully mixed:~$k_{ij} =1$ for $i\ne j$. 
One can be more realistic by taking sparser matrices, but we require ``connectedness":~that for every ordered pair $(i, j)$ of agents there is a path $i_0,\ldots, i_L$ (for some $L>0$ that can depend on $(i,j)$) of positive $k_{i_n,i_{n+1}}$ connecting $i=i_0$ to $i_L=j$.  The rate at which the economy equilibrates depends on how well connected the agents are. 
{If $k$ is very sparse, e.g.~agents arranged in a circle with only nearest neighbour encounters, then convergence is expected to be slower than if $k$ is fully connected}.

At an encounter of agents $(i,j)$, in the basic example, they pool their goods and money and redistribute the totals between them with probability distribution proportional to the product of their utilities. {Thus, agents will typically, although not always, redistribute goods and money in ways that are mutually beneficial according to their utility functions.} Methods for {implementing} this are described in Appendix~\ref{app:B}.  
{Note that}, for CD utilities, integer values of the exponents allow ``direct'' simulation; but for flexibility to allow non-integer exponents or other forms of utility function, most of our simulations use the Metropolis-Hastings algorithm.

We suppose the encounters and redistributions {of goods and money between pairs of agents} all to be independent given the current state, so {that} the dynamic is a continuous-time Markov process on the bisimplex $\sum_i g_i = G, \sum_i m_i = M$, $g_i\ge 0, m_i \ge 0$.

Restricting attention for notational simplicity to the case where each agent's utility depends only on its own holdings, the process has a stationary probability distribution with density 
\begin{equation}
\rho=\frac{1}{Z} \prod_i u_i(g_i,m_i),
\end{equation}
restricted to the bisimplex,
where $Z(M,G)$ is a normalisation factor.  This can be seen by noting that the process is reversible\footnote{This Markov process concept of reversibility is distinct from that for thermodynamics.  An allowed change of state in thermodynamics is reversible if its reverse is also allowed. {Notably, paralleling many cases in physics, the microdynamics of our model is reversible; but the system is not reversible at the macro-level, as a direction is imposed by the non-decrease of entropy.}} with respect to this distribution, i.e.~the flux of probability density from one state in this distribution to another is equal to that for the reverse transition.
Under suitable conditions (including connectedness), it is the only stationary probability distribution and furthermore, any initial probability distribution converges to it (one reference relevant to this is \cite{DGS}; the specific case of all-to-all encounters with just money and $\eta=1/2$ is equivalent to the Kac model of a gas by considering money to be energy and the ergodicity of this is proved in \cite{CCL};
a treatment analogous to that for discrete-time processes with independent transitions in \cite{M} might be possible, but would require extension to continuous-time and to exchange processes).

As mentioned, we can allow different agents to have different utility functions.
We will also allow utility functions that introduce dependency between the {quantities} of goods and money; a simple example is $u(g,m) = (ag+bm)^{\gamma-1}$ for some constants $a,b,\gamma >0$, which makes money and goods {perfect substitutes (in ratio $a:b$)}.  This can be generalised to forms like $u(g,m) = g^{\alpha-1} m^{\eta-1} (ag+bm)^{\gamma-1}$.  Similarly, we can make money and goods complementary by a utility function of the form $u(g,m) = \min(g,m)^{\gamma-1}$ or variants.\footnote{{Substitutes and complements are of considerable importance in economics. To gain an intuition for the distinction,  substitute goods can replace each other (e.g., washing powder vs laundry tablets; or, for less precise substitutes, different brands of soft drink) whereas complementary goods have more value when used together (e.g., to take an extreme example, left and right matching shoes, neither of which has much value on its own).}} We can also make the utility function for one agent depend on the holdings of other agents as well as its own holdings, to incorporate social {preferences}, {in the sense mentioned above}.

Thermal macroeconomics requires the notion of putting two economies in financial contact.  This allows money transfer without goods. 
One interpretation is that some agents have an affiliated agent in the other economy to whom they can send money; another interpretation is that the same individual may play the role of an agent in more than one economy, and may choose to move money from an account in one economy to an account in the other.
To simulate the flow of money, we suppose that the encounter rate matrix $k$ is extended to the union of the two economies (with possibly many zeroes:~only a few agents in each economy need to interact with agents in the other economy) but that in encounters between agents in different economies only their money is pooled and redistributed, with probability distribution proportional to the product of their utilities ({where these utilities incorporate} their fixed {quantities} of goods).  

The theory also requires the notion of putting an economy into contact with a special agent, {the ``trader'', who is} external to the economy, and who offers to buy or sell goods for money at a given price $\mu$. To simulate such interactions, we suppose agent $i$ has encounter rate $K_i$ with the trader ($K_i$ can be zero for many agents) and at an encounter, the agent's initial {quantities} $m^0,g^0$ of money and goods are changed to $m,g$ with distribution proportional to its utility function on the ``price line'' \begin{equation}m+\mu g = m^0+\mu g^0,
\end{equation}
{which represents possible changes of the agent's holdings when buying or selling at the trader's prices}. The trader is assumed not to care about its own {holdings} of goods or money, {so there is no factor from the trader in the formula for the transition probability}.

Lastly, we allow trading contact between economies, in which both goods and money can flow.  We extend the encounter rate matrix to the union of the sets of agents (with possibly many zeroes) and use the same rule for the redistribution of goods and money.

Many variants of the basic set-up are possible.  For example, agents might restrict the outcomes to those in which they don't give away both goods and money.  Then the distribution for the goods and money after an encounter is restricted to the union of the rectangle
$0\le g_i\le g_i^0, m_i^0\le m_i\le m_i^0+m_j^0$ and the one with $i$ and $j$ interchanged. By reversibility, the same stationary distribution holds.  One could even break the symmetry between the agents in the encounter, restricting to just the first rectangle (so $i$ sells, $j$ buys), as long as $k$ is chosen symmetric to preserve reversibility.\footnote{Actually, it might not be necessary to preserve reversibility; reversibility is useful to know the stationary distribution explicitly but that is not essential.  On the other hand, non-reversible dynamics  might lead to failure of transitivity of financial equilibrium, which would violate an assumption of thermal macroeconomics.}
A difficulty with variants in which an agent does not simultaneously give away money and goods is {that is not clear} how to adapt them to financial contact, in which necessarily one agent will give away money and receive nothing in return, hence we don't use it here.  

Another variant is that agents might offer only a fraction of their goods and money at each encounter. 
For example, {suppose that} agent $i$ makes a fraction $f_i$ of its goods and money available at each encounter. {Then} the probability density for the outcome is the normalised restriction of the product of the utilities on the bisimplex to the region with $(1-f_i)g^0_i \le g_i \le g^0_i + f_j g^0_j$ and similarly for $m_i, g_j, m_j$.

Further variants incorporating stylized insights from behavioural science will be described in Section~\ref{sec:beh}. Note that more realistic (and perhaps heterogeneous) models of agents' behaviour will typically be too complex to {analyse by hand}  at the micro-level; the hope is that the aggregate behaviour of economies containing models with such agents will nonetheless be predictable using thermal macroeconomics.

\section{Convergence to equilibrium}

The first test we perform is to confirm that economies of the above forms go to equilibrium.  This is not a test of our thermal macroeconomic theory, but a verification that our simulated economies satisfy the first axiom A0 for the theory in \cite{CM}.
In principle we should check all the axioms of \cite{CM} apply but we leave {this somewhat laborious task} for future work. By testing the predictions of the model as a whole, we are able to validate the usefulness of the approach more directly.

The definition of going to equilibrium is that, given the amounts of the relevant conserved quantities, there is a unique stationary distribution, and from an arbitrary initial probability distribution, the probability distribution as time goes to infinity converges in a suitable metric to {that} distribution.

To test this by simulations requires some discussion, however (as for any Markov chain Monte Carlo simulation), because the simulation produces a sample trajectory, not the evolution of a probability distribution.  
We {test for convergence to a stationary distribution in two ways, using time-averages and aggregate quantities.}

\subsection{Time-averages}
A standard test is to check that the time-average of a selected function along a simulated trajectory converges to the mean\footnote{This time-average is generally called the ``expectation'', but that is somewhat misleading terminology because there is no reason to expect to obtain this value for a single sample.} of the function in the stationary distribution.  A slightly more sophisticated approach is to check that the empirical distribution up to time $t$ converges to the stationary distribution (in a suitable metric).

The convergence of both of these is typically like $t^{-1/2}$, which is slow (though see the comment in \cite{M} about how to view time-averages as producing exponential accuracy for the probability of being within a given distance of the stationary distribution).  Also, time-averages do not distinguish asymptotically stationary behaviour from asymptotically periodic behaviour or any other behaviour with a well-defined time-average.
Nonetheless they provide a useful check.

So, we examine how the time-average of the {quantity} of money held by a particular agent evolves.  We use the simplified quantification of time mentioned in Appendix~\ref{app:A}, namely the number of encounters divided by the sum of the matrix elements of $k$.

\begin{figure}[htbp]
    \centering
\subfigure[\textbf{}]{
\includegraphics[height=1.5in]{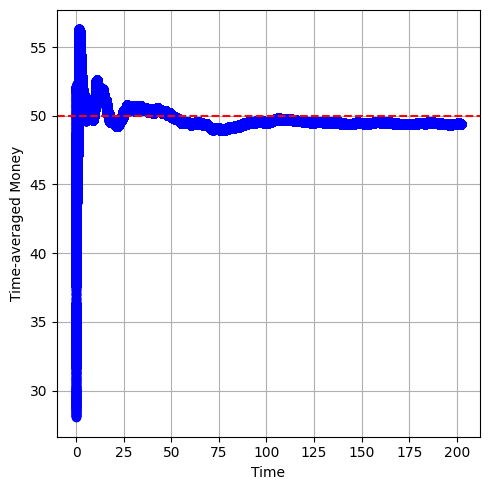}
\label{a1}
}
\quad
\subfigure[\textbf{}]{
\includegraphics[height=1.5in]{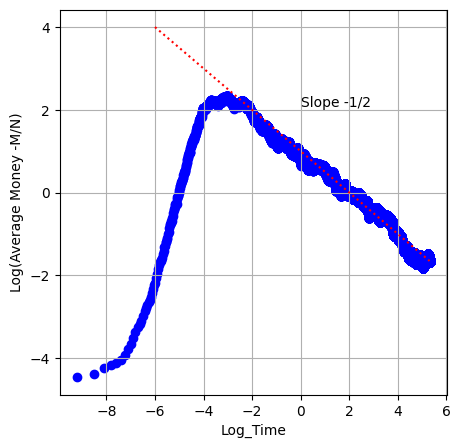}
\label{a2}
}
\quad
\subfigure[\textbf{}]{
\includegraphics[height=1.5in]{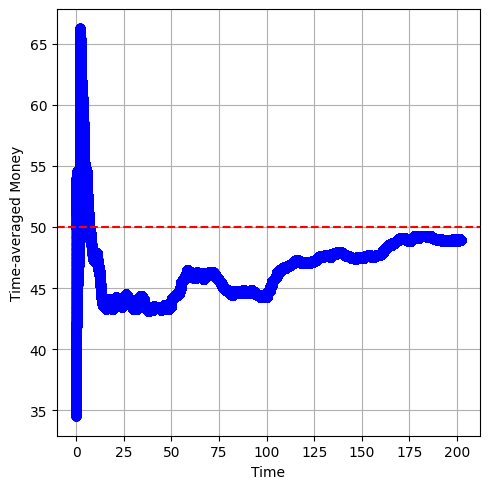}
\label{b1}
}
\quad
\subfigure[\textbf{}]{
\includegraphics[height=1.5in]{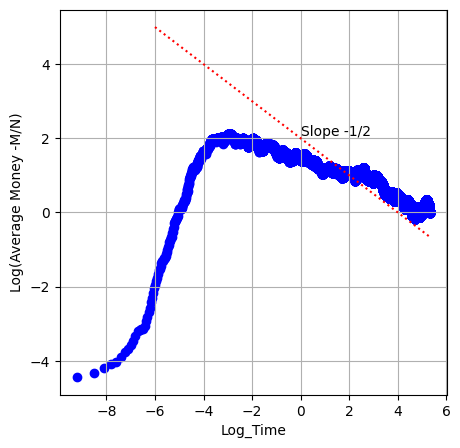}
\label{b2}

}
\quad
\subfigure[\textbf{}]{
\includegraphics[height=1.5in]{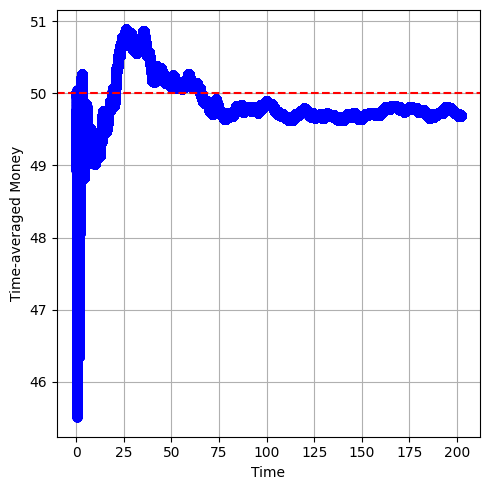}
\label{c1}

}
\quad
\subfigure[\textbf{}]{
\includegraphics[height=1.5in]{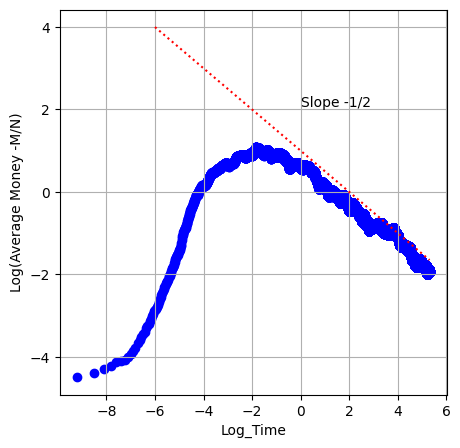}
\label{c2}

}
\caption{(a) Time-averaged money $\bar{m}_i(t)$ for a chosen agent $i$ in a Cobb-Douglas economy of $N=100$ agents with $\alpha=2, \eta=3$, total money $M=5000$ and goods $G=5000$.  The simulation was initialised with equal money and goods for each agent.  {It shows slow convergence towards the expected value $M/N = 50$, modulo fluctuations.} 
(b) Average of $\log |\bar{m}_i(t)- M/N|$ over 100 samples, 
demonstrating the $t^{-1/2}$ 
convergence (slope $-1/2$) after an initial transient.
(c) Time-averaged money $\bar{m}_i(t)$ for a chosen agent $i$ in the same CD economy but each agent only trades with two neighbours.
(d) Average of $\log |\bar{m}_i(t)- M/N|$ over 100 samples for CD economy with 2 neighbours.
(e) Time-averaged money $\bar{m}_i(t)$ for a chosen agent $i$ in the same CD economy but agents only trade $10\%$  of their money in each interaction. (f) Average of $\log |\bar{m}_i(t)- M/N|$ over 100 samples for CD economy with fractional trade.  
    }
    \label{fig:cge1}
\end{figure}

Figure~\ref{fig:cge1}(a,c,e) shows the time-averaged money as a function of time for one agent from a homogeneous Cobb-Douglas economy with $\alpha=2, \eta=2$, under three cases of interaction.  These are:
\begin{enumerate}
\setlength{\parskip}{-1ex}
\item[(i)] the basic case (encounters between all pairs of agents and redistribution between them according to the product of their utilities); 
\item[(ii)] a sparse case ({the agents are arranged in a circle and encounters are only between neighbours}); 
\item[(iii)] a fractional case (agents make only a fraction of their possessions available at each encounter, but still redistribute according to the product of their utilities for the outcome).
\end{enumerate}
The simulations were started from {a} uniform distribution of money and goods over the agents (the exchange of goods is irrelevant in this simulation but we left it in for completeness).
The mean in the stationary distribution is $M/N$ and indeed the figure shows convergence to this value.

To test that the convergence {is} like the predicted $t^{-1/2}$, we made log-log plots in Figure~\ref{fig:cge1}(b,d,f).  For these, we averaged $\log |\bar{m}_i(t)-M/N|$ over 100 samples and employed a cutoff to reduce the potentially large negative fluctuations (due to $\log |m| \to -\infty$ as $m \to 0$).

\begin{figure}[htbp]
    \centering
    \includegraphics[height=1.9in]{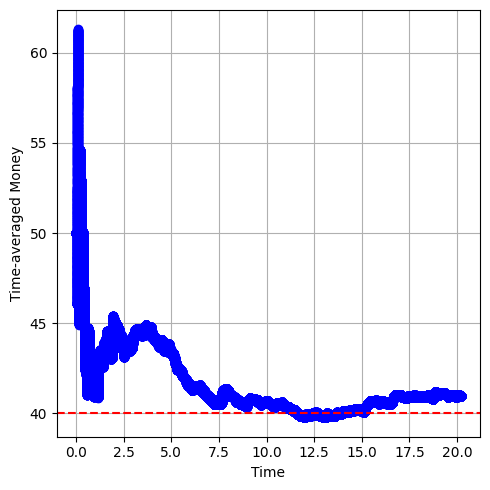}
    \includegraphics[height=1.9in]{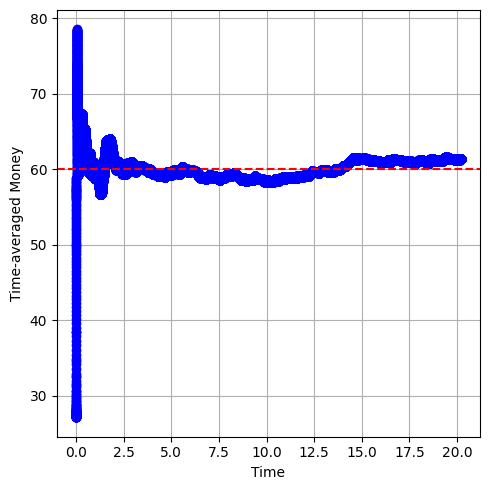}
    \caption{Time-averaged money $\bar{m}_i(t)$ for two chosen agents in a Cobb-Douglas economy of $N=100$ agents with $\alpha=2$; {half the agents} have $\eta_i=3$, {and} the other half $\eta_i=2$; total money $M=5000$ and goods $G=5000$.  The simulation was initialised with equal money and goods for each agent.  {The left-hand figure depicts time-averaged money for an agent with $\eta_i=3$; the right-hand figure depicts results from an agent with $\eta_i = 2$.}}
    \label{fig:cge2}
\end{figure}

Figure~\ref{fig:cge2} shows the time-averaged money for two agents in an inhomogeneous economy, with two types of agent with utility functions with different exponents and the basic interaction.  In this example, the time-averaged money for an agent with exponent $\eta_i$ should converge to $\eta_i T$, with ``temperature'' : 
\begin{equation}
    T = \frac{M}{\sum_j \eta_j},
    \label{eq:temperature}
\end{equation}
(extending the analysis of \cite{CM}, section 12) as illustrated by the figure. 

\subsection{Aggregate quantities}
An alternative approach to testing convergence to equilibrium is to choose a function that is an aggregate in some sense, e.g.~an average over agents, and examine how it behaves along a trajectory.  The advantage of an aggregate is that for a large system its stationary distribution is in general much {less variable} than its range of possible values.  For example, the average money per agent in a subset of size $n$ of a population can be expected to lie within order $n^{-1/2}$ of its mean with large probability (we don't consider the full population because for a single economy the total money is conserved so the average money per agent is constant).  Then the typical behaviour of a simulation from an arbitrary initial condition is a ``burn-in'' period in which the value of the aggregate relaxes from order 1 to within order $n^{-1/2}$ of its stationary mean, followed by fluctuations of this order about the mean.  

A theory of burn-in for Markov processes on large networks, developed by one of the authors (to be written up elsewhere, though for independent update processes rather than exchange processes), provides a quantitative analysis of the approach to equilibrium.\footnote{We note that some authors, e.g.~\cite{G}, suggest that either there is no theory of burn-in and/or that no such theory is necessary. In the present context, though, the theory does exist and produces accurate results.} Thus, observing this burn-in behaviour provides an indirect test for convergence to equilibrium.

With this in mind, we examine how some aggregate functions behave along trajectories.  Rather than taking the aggregate to be an average of some quantity over agents, we look at the fraction of agents with money less than a given value; and we let the given value vary over the whole range, i.e.~we plot the empirical cumulative distribution for money over the population.  The marginal distribution for one agent in a homogeneous population of size $N$ with total money $M$ is a Beta($\eta,(N-1)\eta$) distribution scaled to the range $[0,M]$, i.e.~density 
\begin{equation} \rho(m) \propto m^{\eta-1}(M-m)^{(N-1)\eta - 1}.
\end{equation}
The stationary distribution for the empirical cumulative {function} is within order $N^{-1/2}$ of the cumulative {function} for the Beta with large probability.  For $N$ large, one can approximate the Beta distribution very well by a Gamma distribution with parameter $\eta$ and mean $\bar{m} = M/N$, i.e.~density $\rho(m) \propto m^{\eta-1}e^{-\eta m/\bar{m}}$, so we compare the empirical distribution with this as a function of time.  

We quantify the difference between two cumulative distributions by the Kolmogorov-Smirnov (KS) statistic, which is simply the maximum absolute value of the difference.  By extrapolation from the case of independent draws from a continuous 1D cumulative distribution, this can be expected to settle to fluctuations of size $N^{-1/2}$.\footnote{More precisely, the KS statistic is distributed asymptotically for large $N$ as $N^{-1/2}B$, where $B$ is the Brownian bridge (the random variable given by the maximum of a standard Brownian motion on $[0,1]$ started at 0 and conditioned on ending at 0).} 
We can alternatively use the Kantorovich-Rubinstein (KR) distance, which is the integral of the absolute value of the difference between the cumulative distribution functions.

\begin{figure}[tb]
    \centering
    \includegraphics[height=1.9in]{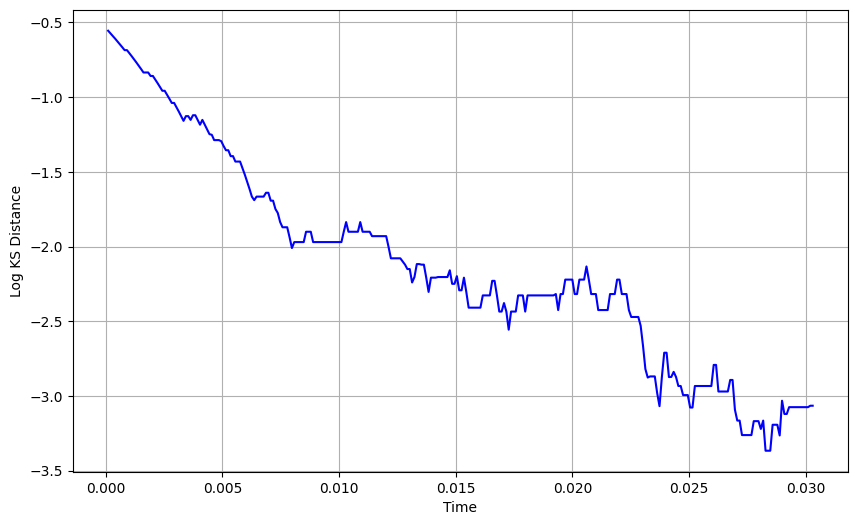}
    \includegraphics[height=1.9in]{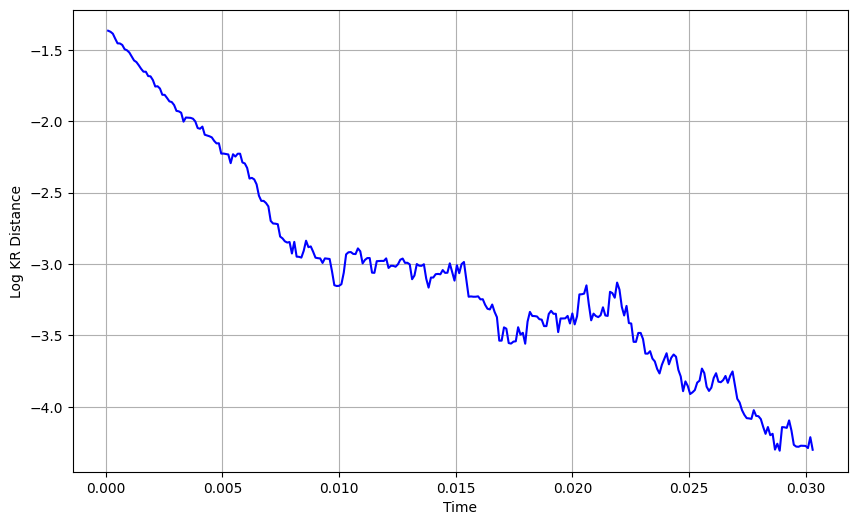}
    \caption{Logarithms of the KS and KR statistics against time, for the cumulative distribution of money over a homogeneous population of agents with $\eta=3$.}
    \label{fig:KSKR}
\end{figure}

Figure~\ref{fig:KSKR} shows how the KS and KR statistics behave as functions of time for one realisation of a homogeneous population, starting from equidistribution.  
The log-linear plots demonstrate an exponential burn-in period followed by the expected size of fluctuations about the ideal.  
The burn-in rate {is} about $1/0.01$ per unit of time, which is reasonable because we were using the fully mixed encounter matrix, so we would expect a rate {of }roughly $N$, which was $100$.

\begin{figure}[htb]
    \centering
\includegraphics[height=1.9in]{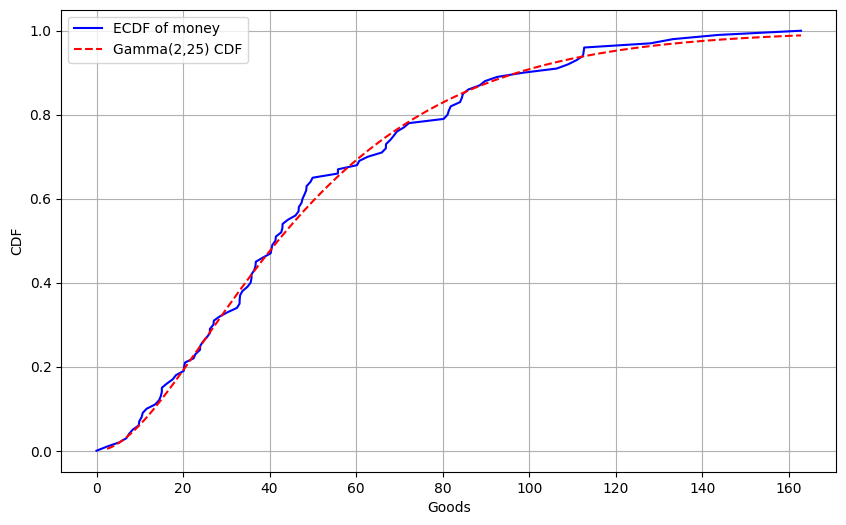}
\includegraphics[height=1.95in]{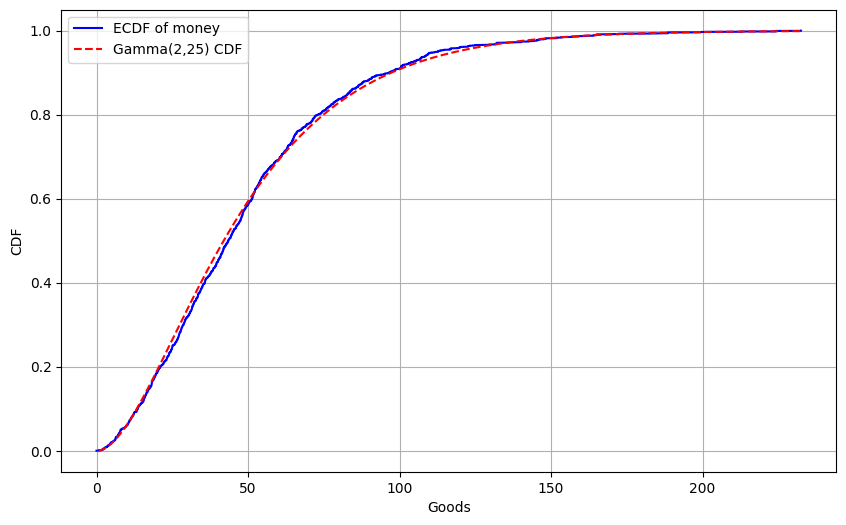}
    \caption{The first graph shows the wealth distribution for a {single} CD economy with 100 agents and $\eta=3$, after time 1, compared with the Gamma approximation for the stationary distribution.  The second graph shows the average of the wealth distribution over 1000 samples.}
    \label{fig:wealth}
\end{figure}

Figure~\ref{fig:wealth} shows the empirical cumulative distribution at some time after the burn-in period, compared to the ideal cumulative distribution, for $N=100$.  Statistical deviations are visible. The second panel shows an average over 1000 samples and the difference from the ideal is {barely visible} to the eye.

For an inhomogeneous population of agents with different Cobb-Douglas exponents, some modifications would be required.  For an agent $i$ in a heterogeneous economy, the marginal of the stationary distribution for its money is Beta$(\eta_i, \sum_{j\ne i} \eta_j)$.

\section{Fluctuations}

As we have seen in the previous section,
the amount of money owned by an agent, or a subset of agents, fluctuates in time about its mean.  
{Thermal macroeconomics says nothing about fluctuations, because it is a theory for macroscopic quantities in the idealised case of infinite systems. {Nonetheless,} for practical purposes it is important to understand the typical size of fluctuations.}

For our stochastic models, we can predict the variance of fluctuations.
In particular, taking a relatively small subset $S$ of agents and denoting the sum of their money by $m_S$, we expect $\Var\, m_S \approx C_S T^2$, where the ``money capacity''  $C_S = \sum_{j\in S} \eta_j$ {by analogy with heat capacity in thermodynamics}, and $T$ is the temperature, as explained in \cite{CM}. 

We tested this for a CD economy where half of the agents have $\eta=2$ and the other half have $\eta=3$.  The total number of agents was $N=100$ and total amount of money $M=5000$. Thus, from (\ref{eq:temperature}) the temperature $T = 20$.  For subset $S$, we chose random subsets of size 3, so they have money capacity $C_S$ equal to 6,7,8 or 9.  The variance of the money in $S$ was estimated by taking $n=1000$ samples and 
computing the average of the squared deviations from the sample mean\footnote{{In principle, one should divide by $n-1$ instead of $n$, but with $n=1000$ it makes negligible difference.}}.  

The results are shown in Table~\ref{tab:variance} {and show reasonable agreement with our theoretical predictions}.  {The systematic under-estimate can be attributed to the ratio of the size of $S$ to the total population $(3\%)$. An exact formula (\ref{eq:varf}) for the variance is given in the next section. For the case $C_S = 6$, for example, it predicts $\mbox{Var } m_S = 2333.1$, which agrees well.} 

\begin{table}[htb]
    \centering
\includegraphics[height=2.2in]{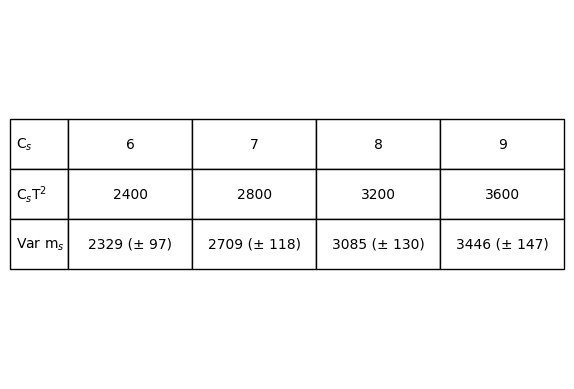}
    \caption{Variance of money in chosen subsets of agents. Half of the agents have $\eta_1=2$ and half have $\eta_2=3$.  There were $N=100$ agents, with total money $M=5000$, so from \eqref{eq:temperature} the temperature was $T=20$. Subsets of three randomly chosen agents were selected, so they have money capacity $C_s = 6,7,8$ or $9$, according to their combination of $\eta$ values.}
    \label{tab:variance}
\end{table}

\section{Effects of financial contact}

The next test is to put two economies in financial contact and monitor their temperatures.  The theory of \cite{CM} {gives a concept of temperature to an economy and} predicts, by analogy with thermodynamics, that nett money should flow from the hotter to the cooler and their temperatures should converge together (modulo fluctuations). This is consistent with an interpretation of temperature as the inverse value of money:~money flowing from hotter to cooler thus corresponds to money flowing to wherever its value is greater \cite{CM}. 

For our {CD} economies, the equilibrium temperature from \eqref{eq:temperature} is exactly 
\begin{equation}
T= M/C
\label{eq:temp}
\end{equation}
with $C=\sum_i \eta_i$, {which {is} its money capacity {as introduced above}} \cite{CM}.  So we use $T^A = M^A/C^A$ for the temperature of a {CD} economy $A$ where $M^A$ is the amount of money in {the economy} and $C^A$ its money capacity.  

For two {CD} economies $A$ and $B$ in financial 
contact, the fraction \begin{equation}
f^A = \frac{M^A}{M^A+M^B}
\end{equation}
of the total money that is in $A$ at equilibrium is theoretically distributed as Beta($C^A,C^B$).  This has mean $C^A/(C^A+C^B)$ and variance 
\begin{equation}
\mbox{var}(f^A) = \frac{C^A C^B}{(C^A+C^B)^2(C^A+C^B+1)}.
\label{eq:varf}
\end{equation}
In particular, for $C^A, C^B$ large, {the distribution} is tightly {peaked} around its mean. {The thermal macroeconomic theory is expected to apply in the thermodynamic limit where the number $N$ of agents goes to infinity with $M$ and $C$ proportional to $N$.  Then we see that the variance of $f^A$ goes to zero like $1/N$.}

\begin{figure}[h]
    \centering
\includegraphics[height=2.0in]{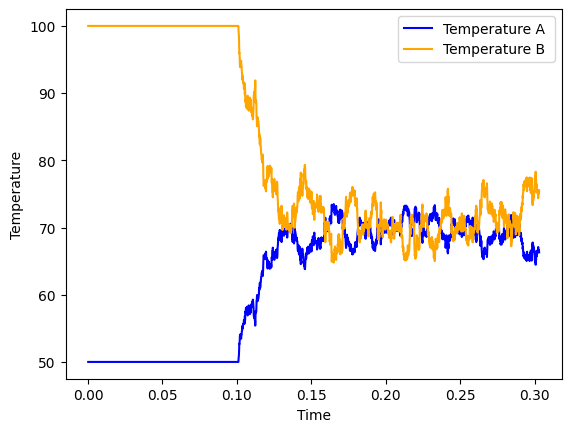}
\includegraphics[height=2.0in]{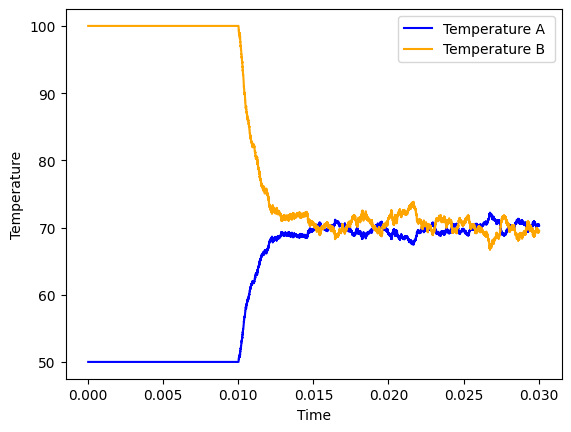}
\caption{{(a) Temperatures of two economies put into financial contact at time $0.1$, as functions of time. The parameters are $\eta_A=3$, $\eta_B = 2$, $N=100$ for each, and initial $M_A = 15000$, $M_B = 20000$.}
(b) Temperatures of the same two economies put into financial contact at time 0.01, using the trick of implementing only $10\%$ of the change at each encounter.}
\label{fig:equiltemp}
\end{figure}

Figure~\ref{fig:equiltemp}(a) shows the result of putting two CD economies into financial contact.  They each have $N=100$ agents.  Economy A has $\eta_A = 3$ and initial money $15000$, {and} hence temperature $50$. 
Economy B has $\eta_B = 2$ and initial money $20000$, {and }hence temperature $100$. The initial period before financial contact is used to ensure that each economy has equilibrated internally.

We see that money {does indeed} flow to make the temperatures equilibrate, modulo fluctuations.  We also see that the direction of nett money flow is from hotter to cooler, because the amount of money in a CD economy is proportional to its temperature.

We can again ask whether the fluctuations are of the expected size.  Using (\ref{eq:temp}) and (\ref{eq:varf}), the standard deviation of $T^B$ should be $3.83$.   The results in the figure look approximately of this size, and measurement of the standard deviation from 500,000 samples after settling to equilibrium yielded a value of $3.82$. 

To reduce the fluctuations, we could increase $N$.  For example, $N=1000$ with the same total amount of money shared between them should reduce the standard deviation of $T^B$ by a factor of $\sqrt{10}$ to $1.21$.  But going to $N=1000$ stretches laptop capacity for time and memory.
Instead, we employ a trick that achieves an effective increase in $N$ without increasing the number of agents.

The trick is to implement only a fraction, say $1/10$ for illustration, of the change computed at each encounter.
This simulates each agent consisting of 10 sub-agents, provided that their utility functions are homogeneous.  A utility function $u$ (of two variables for illustration) is {\em homogeneous} if there is a $\tau$ such that for all $\lambda>0$,
\begin{equation}
u(\lambda g, \lambda m) = \lambda^\tau u(g,m).
\end{equation}
This holds for CD economies and several (though not all) of the variants we shall consider.  Implementing $1/10$ of the change at an encounter between two such agents is equivalent to implementing a normal encounter between two of their sub-agents, if the possessions are considered to be distributed equally among the sub-agents.  
A weakness in this argument is that for a genuine simulation of 10 times the number of agents, the possessions in a group of 10 are not distributed uniformly.  

A more justified strategy, at least in the case of a CD economy, is to select a Beta($\eta,9\eta$) fraction of the money in the group for redistribution, instead of $1/10$.  
This gives an exact simulation in the limit that we consider sub-agents within a group to encounter each other infinitely fast.  Nevertheless, Beta($\eta,9\eta$) is peaked around its mean of $1/10$ (from (\ref{eq:varf}), the standard deviation is $0.3/\sqrt{10\eta+1}$), so we think the effect of using $1/10$ will be close to the exact simulation. Thus, we did not implement the exact simulation (furthermore, if the two economies have different values of $\eta$, further modification is required).
Another minor comment is that implementing only a fraction of the result of an encounter also slows down the process, when time is measured by number of encounters, so to avoid this effect one should rescale time by the same factor.

The result of the trick is to multiply the effective number of agents by 10 and hence reduce fluctuations by a factor of $\sqrt{10}$.
This is borne out in Figure~\ref{fig:equiltemp}(b), where the trick is implemented on the same pair of economies. Indeed the measured size of the standard deviation for $T^B$ from 500,000 samples was $1.15$, close to the prediction of $1.21$.  Note that the timescale for convergence is also stretched by a factor $10$, as expected.

\section{Price}

Next, we test the prediction that {each good in an economy has a ``market price'', such that} on putting the economy into contact with a trader offering a fixed price for a good, the economy will {in aggregate} buy from the trader if the trader's price is below the economy's market price, or {in aggregate sell to the trader if} the trader's price is above the market price.

\begin{figure}[htb]
    \centering
    \includegraphics[height=2.2in]{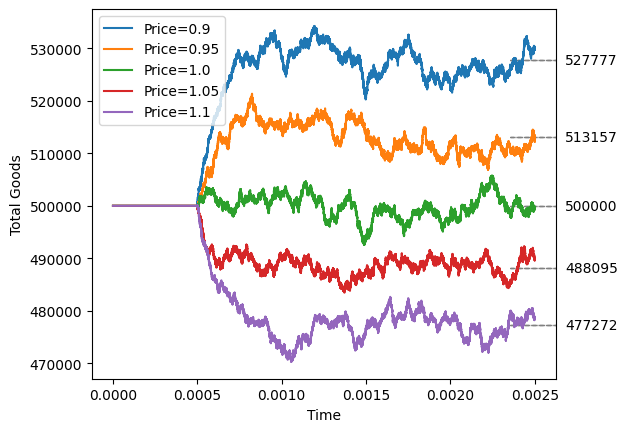}
    \caption{Changes in {quantities} of goods in a CD economy on connecting to a trader offering the indicated prices.  The economy had $\alpha=2$, $\eta=2$, $N=10,000$, $M=G=500,000$, so the initial market price in the economy was 1.0.  The predicted equilibrium {quantities} of goods in the economy at these prices are indicated on the right.}
    \label{fig:goodsflow}
\end{figure}

For a homogeneous CD economy, the market price of a good  \cite{CM} is
\begin{equation}
\mu = \frac{\alpha M}{\eta G}.
\end{equation}
Figure~\ref{fig:goodsflow} shows the effects {on the quantity $G$ of a good} when a homogeneous CD economy {is connected} to a trader offering prices in a range about the initial market price.
The direction of nett flow of goods
agrees with the prediction.
Note that the market price for the good changes with the {quantity} of goods in the economy.
So we see also that the {quantity} of goods in the economy settles to a new level at which the market price equals the trader's price. {To confirm this, we compute the expected {quantity} of goods:~trading happens on the price line $M+\mu G = M_0 + \mu G_0$, where $M_0, G_0$ are the initial money and goods.  $M, G$ should settle to make $\frac{\alpha M}{\eta G} = \mu$.  The solution of these two equations is 
\begin{equation}
G = \frac{M_0+\mu G_0}{(1+\eta/\alpha)\mu}.
\end{equation}
The values of $G$ for the chosen prices are indicated in the figure. {As expected, when the trader offers a low price, the economy buys goods from the trader so that $G$ increases; and conversely, when the trader offers a high price, the economy sells to the trader so that $G$ decreases. As the figure indicates, the equilibrium values of $G$ as a function of the price offered by the trader agree closely with theory.}}

For a homogeneous CD economy, both $M$ and $G$ are easily measured. So a way to measure the price of a good in a general exchange economy is to connect a smaller CD economy to it, allowing flow of money and goods, and measure price there by time-averaging $\mu$ in the CD economy.  This will be elaborated in the next section.

\section{CD meters}
\label{sec:CDmeter}

The results of the previous two sections for CD economies motivate the use of CD economies as meters for measuring the temperature of a general simulated economy, or the price of a good in it.  

The idea is to attach a CD economy to the general one, let it equilibrate, and measure the temperature, value or price in the CD economy by the formulae there.
To recall, the formulae are
\begin{equation} T = \frac{M}{C}, \quad \mu = \frac{C_G M}{C G},
\end{equation}
where $M,G$ are the money and goods in the CD economy, $C = \sum_i \eta_i$ is the money capacity, and $C_G = \sum_i \alpha_i$ can be called the goods capacity.  For use as a meter, {it is particularly convenient if the} CD economy is homogeneous, meaning all agents have the same $\alpha$ and all have the same $\eta$.  Then $C = N\eta$ and $C_G = N\alpha$.

There are two issues to bear in mind.  Firstly, the {quantities} of money and goods in the CD economy  fluctuate, so it is best to time-average them over some period.  Secondly, attaching a CD meter in general changes the average {quantities} of money and goods in the target economy.  A trick to avoid the latter, however, is to artificially restore the {quantities} of money and goods in the target economy after each encounter with the CD economy. 
Equivalently, at each encounter between agents in the two economies, {we can} update the CD agent's holdings according to the rules but leave the target economy agent's holdings unchanged. 

The same idea can be used to measure the (marginal) value of a good in a general economy.  This is defined as the derivative of the entropy with respect to the {quantity} of the good.
The value of a good in a CD economy is 
\begin{equation}
\nu = \frac{C_G}{G}.
\end{equation}

\section{Trade}

{We have so far dealt with trade between an economy and the trader. But two} economies can {of course also} be put into trading contact {with each other}, i.e.~allowing exchange of goods and money between agents in different economies, using the agents' utilities, {just as in exchanges within an economy}.  A prediction of the theory is that, {just as in classical thermodynamics,} total entropy can not decrease.

Thus, we monitor the entropies of the two economies.  For the basic case of a {homogeneous} CD economy, the entropy $S = N \log G^\alpha M^\eta$ \cite{CM}.  For a heterogeneous CD economy, $S = C_G \log G + C \log M$, where $C_G = \sum_i \alpha_i$, $C= \sum_i \eta_i$.
We {can thus now check by simulation} that the total entropy does not decrease (modulo fluctuations because we are not in the thermodynamic limit).

\begin{figure}[h]
    \centering
    \includegraphics[height=2in]{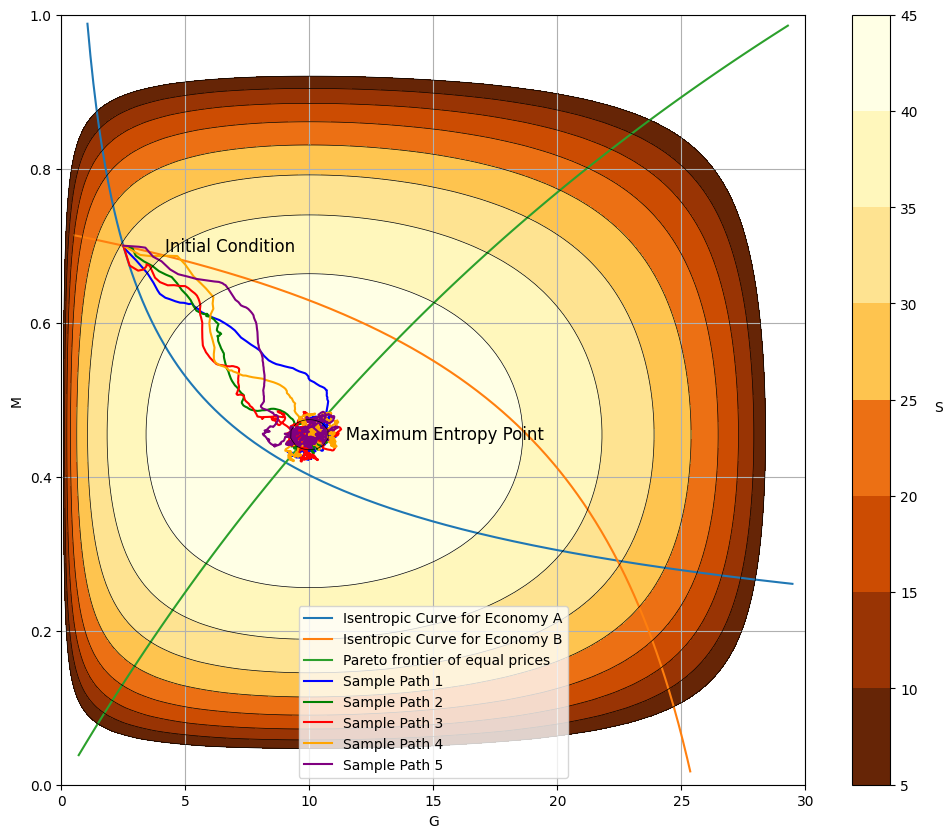}
    \includegraphics[height=2in]{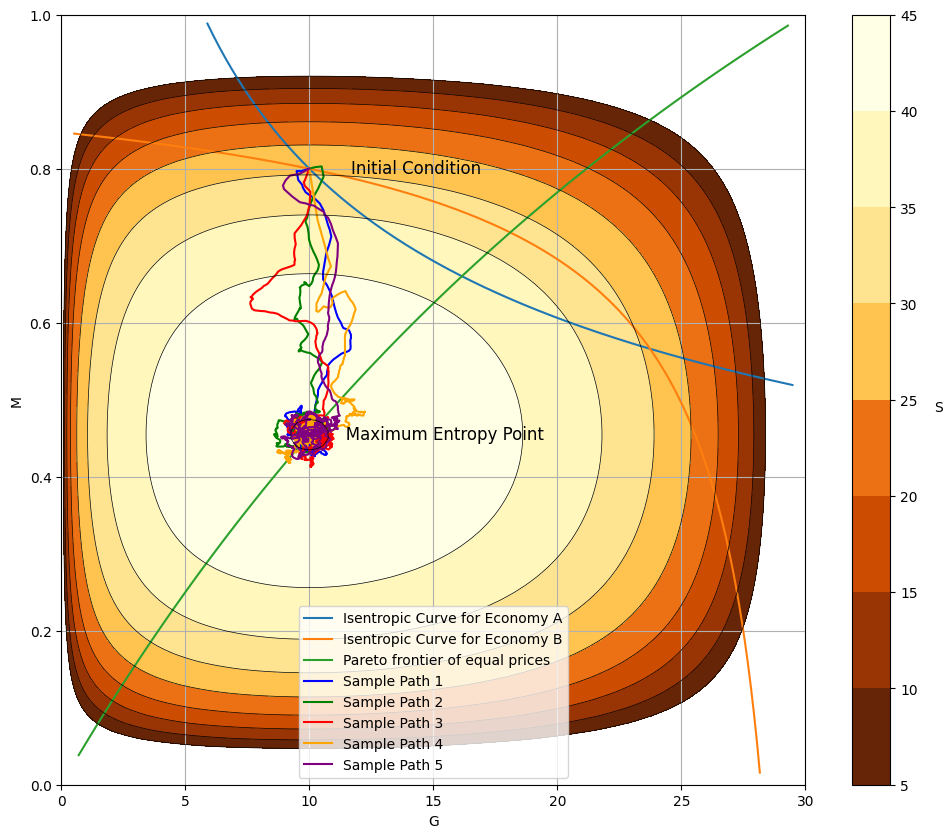}
\caption{Sample paths in the space of $(G_1,M_1)$ for two CD economies put in trading contact from two different initial endowments. They have $\eta_1 = \frac{5}{2}$, $\eta_2 = \frac{3}{2}$, $\alpha_1 = \alpha_2=1$, $N_1 = 10$, $N_2 = 20$ and total money $M=1.0$, goods $G=30$.  Also showing contours of constant total entropy, the curves for the initial entropies of economies 1 and 2 (in blue and orange), and the curve of equal prices.
(a) Initial endowment $M_1 = 0.7$, $G_1 = 2.5$, $M_2 = 0.3$, $G_2 = 27.5$.  (b) 
Initial endowment $M_1 = 0.8$, $G_1 = 10$, $M_2 = 0.2$, $G_2 = 20$. {Note that, as expected by theory, these paths tend, modulo fluctuations, to climb the entropy gradient and settle at the point of maximum entropy.}}
    \label{fig:samplepaths}
\end{figure}

Figure~\ref{fig:samplepaths} shows some sample paths for two economies put in trading contact, from two choices of initial endowments.
It confirms that when viewed on a macroscopic scale, trading never leads to decrease of total entropy. 
It shows that trading leads to a probability distribution peaked around the maximum entropy state.

An interesting feature of the second case is that although the total entropy increases, the entropy of economy $1$ actually decreases. 

These observations have an interesting economic interpretation, if we view entropy as a measure of the ``aggregate utility'' of an economy (see \cite{CM} for motivation for, and limitations of, this interpretation). Then, the second law implies that aggregate utility of a system cannot decrease through trading within or between economies (aligning with arguments for markets and free trade in classical economics dating back to Adam Smith and David Ricardo). But the second case above illustrates that trade between economies need not necessarily be mutually beneficial.

\section{The Carnot cycle}

The next test we describe is to
demonstrate that a trader can make money out of a temperature difference between two economies. 
The trader can do this by trading goods with a ``ship'' economy that it puts alternately in financial contact with the two ``mainland'' economies, employing an analogy with the Carnot cycle in classical thermodynamics, which extracts heat from a physical temperature difference in order to do work. 

In our simulations, we model the mainland and 
ship economies as CD econo\-mies.
We start with the mainland economies at temperatures $T_H = 26, T_C = 20$ ($H$ for hot, $C$ for cold).  The ship economy starts at a temperature $T_S = T_C$, the trader buys goods from it at {$3\%$} above market price on the ship, which makes its money and hence temperature rise. We call this an ``isentropic'' phase because trade at exactly market price would conserve entropy on the ship.  {The trader's price is updated after a time corresponding to $N_S$ encounters with the trader.}  When $T_S$ reaches $T_H$, the trader puts the ship into financial contact with the hot economy.  The trader then starts selling goods to the ship at {$3\%$} below market price. {The aim of the financial contact is to keep the temperature on the ship close to that on the mainland, so we call this an ``isothermal'' phase.}  The trader's price is updated after each $10,000 
\,N_S$ encounters. {The reason for updating the price less frequently than in the isentropic phase is that otherwise with the financial contact we find that an instability appears, in which the temperatures of the ship and mainland diverge (it would be interesting to explain this instability).}
As the {quantities of the} goods on the ship increase, money flows from the hot economy to the ship to maintain the temperature. 
Because the hot economy is not infinite, this money-flow out of the hot economy will make the temperature $T_H$ decrease, but to simulate the ideal of an infinite hot economy, we can artificially add money to the hot economy to keep its {total amount of} money, and hence its temperature, constant. 

When the ship economy reaches a certain {quantity} of goods, the trader disconnects the ship from the hot mainland and continues to sell goods to the ship at {$3\%$} below market price and $T_S$ decreases.  When $T_S$ reaches $T_C$, the trader puts the ship into financial contact with the cold mainland and reverts to buying goods from the ship.  Money flows from the ship to the cold mainland in the process, but less than in the hot phase {with the resulting difference in money accruing as profit that can be extracted by the trader}.  The trader disconnects the ship from the cold mainland when the goods decrease to the initial {quantity}, and the cycle restarts. 

\begin{figure}[htb]
    \centering
\includegraphics[height=1.4in]{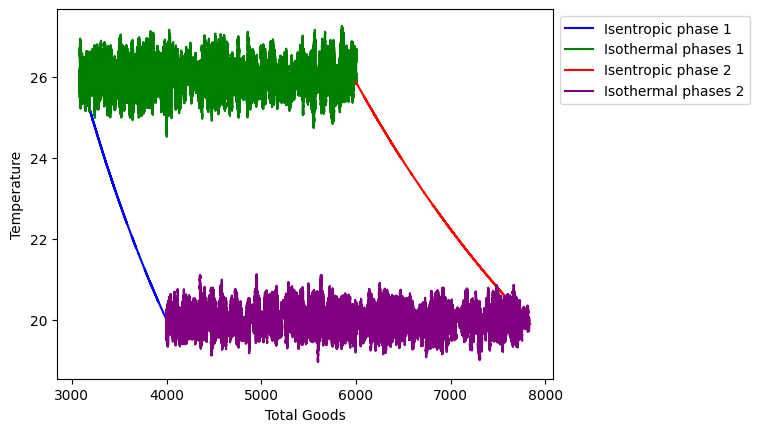} \includegraphics[height=1.4in]{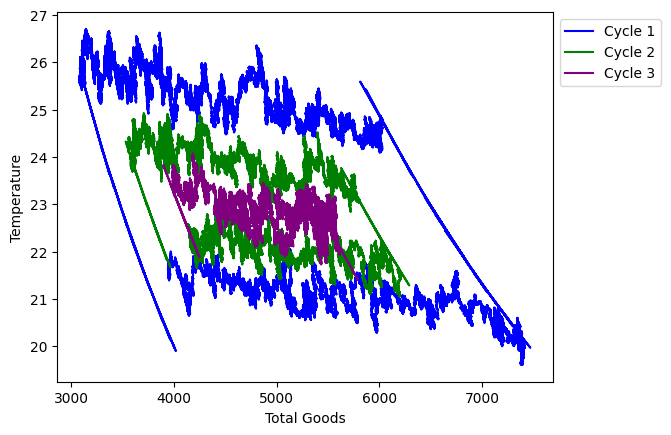}
\caption{(a) One turn of a Carnot cycle, with $T_H$ and $T_C$ maintained at 26 and 20, respectively, showing the {quantity} of goods and temperature on the ship economy (the nett motion is clockwise). For all three economies,
$\eta=2$, $\alpha=2$ and $N=100$. The initial conditions are $M_H$=5200, $G_H$=5200, $M_C$=4000, $G_C$=4000, $M_S$=4000, $G_S$=4000. The trader switches between isothermal and isentropic processes when $G_S$ becomes larger than 6000 or lower than 4000. \\
(b) Several turns of a Carnot cycle, without artificial maintenance of mainland temperatures, showing how the temperatures on the mainlands approach each other (the nett motion is clockwise). For all three economies, $\eta=2$ and $\alpha=2$. 
The ship economy has $N=100$, the hot and cold economies have $N=1000$.
The initial conditions are $M_H$=52000, $G_H$=52000, $M_C$=40000, $G_C$=40000, $M_S$=4000, $G_S$=4000. The trader switches from isothermal to isentropic process when $G_S$ becomes larger than [6000, 5800, 5600] successively, or lower than [4000, 4200, 4400], and switches from isentropic to isothermal when the $T_{S}$ becomes the same as $T_H$ or $T_C$.}
    \label{fig:Carnot}
\end{figure}

\begin{figure}[!h]
    \centering
    \includegraphics[height=2in]{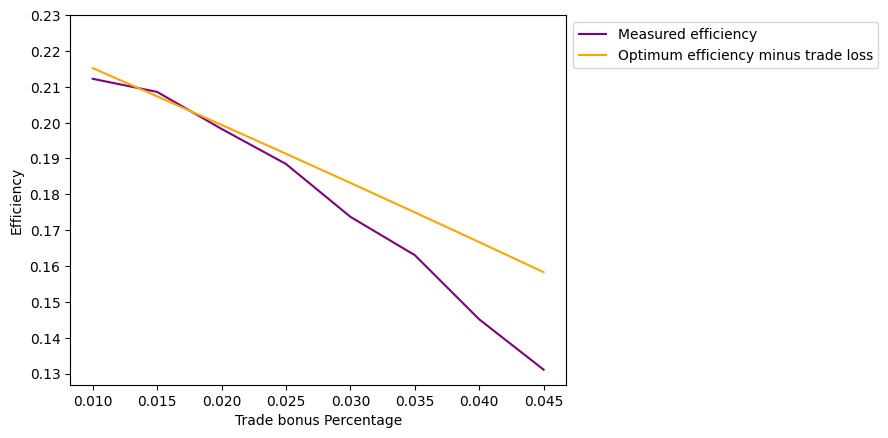}
    \caption{Average efficiency over 10 realisations of the Carnot cycle for various values of the deviation of the trader's price from market price, compared with the theoretical maximum efficiency minus the loss due to the deviation.}
    \label{fig:efficiency}
\end{figure}

Figure~\ref{fig:Carnot}(a) 
demonstrates a Carnot cycle with the artificial maintenance of $T_H$ and $T_C$ described above (at $T_H=26, T_C=20$).  The nett flow of money from the hot economy is $\odot 3334$, and the nett money flow to the cold economy is $\odot 2778$.  The trader makes the difference, $\odot 555$, and so the efficiency of the cycle is $555/3334 \approx 0.167$.
According to thermal macroeconomic theory, the maximum efficiency of a cycle operating between temperatures $T_H$ and $T_C$ is $1-T_C/T_H \approx 0.231$.  
{We see that this simulation 
satisfies the efficiency bound.  The efficiency can be increased by trading at a smaller difference from market price in the isentropic phases.  This is illustrated in Figure~\ref{fig:efficiency}. Efficiency can probably also be improved by taking the isothermal phases more slowly.}


To illustrate the effect of finiteness of the mainlands, we also simulate three turns of a Carnot cycle without the artificial replenishment of money in the mainlands.  In the process, the temperatures of the mainlands move {closer} together, so we also change the thresholds for the {quantities} of goods on the ship for disconnection at each turn, producing a decreasing spiral, 
see Figure~\ref{fig:Carnot}(b).

\section{The Carnot cycle in reverse}

The Carnot cycle allows the trader to extract money efficiently from a temperature difference between two economies. The reverse Carnot cycle does the opposite:~by putting in a relatively small amount of money, a trader can {efficiently} {move money from a colder to a hotter economy, or make a temperature difference between two economies that start at the same temperature, or enhance the temperature difference between two economies.  {This process is} analogous to a heat pump or refrigerator {in thermodynamics}.}  

\begin{figure}[htb]
    \centering
\includegraphics[height=1.4in]{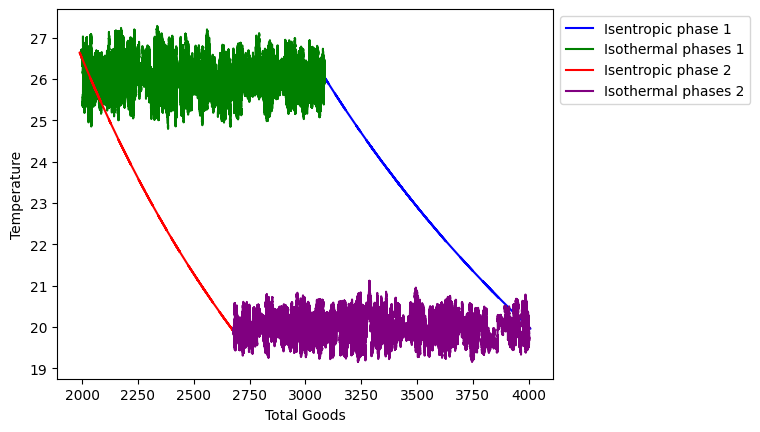} 
\includegraphics[height=1.4in]{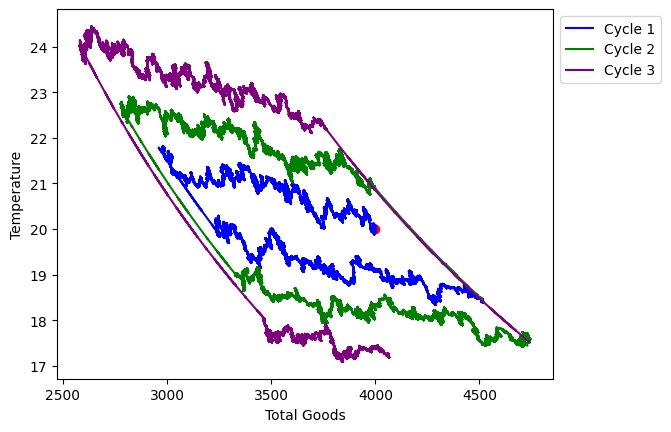}
\caption{(a) One turn of a reverse Carnot cycle, with artificial maintenance of the mainland temperatures (at $T_H=26,T_C=20$), showing the {quantity} of goods and temperature on the ship economy (the nett motion is anti-clockwise).
For all three economies, $\eta=2$, $\alpha=2$, $N=100$, with initial $M=4000$, $G=4000$ on the ship.  The system transfers between isothermal and isentropic processes when $G_S$ becomes larger than 4000 or lower than 2000. \\
(b) Three turns of a reverse Carnot cycle, without artificial maintenance of the mainland temperatures, showing how the temperatures on the mainlands diverge from the initial temperature $T=20$. For all three economies, $\eta=2$, $\alpha=2$. For the ship economy $N=100$ and for the cold and hot economies $N=500$,  and the ship economy starts with $M=4000$, $G=4000$ (shown by a red dot).  The hot and cold economies start at the same temperature as the ship economy  but with $M=20,000$, $G=20,000$.  The system transfers from isothermal process to isentropic process when $G_S$ becomes lower than [3000, 2800, 2600] on successive turns or larger than [4500, 4700],
and transfers from isentropic process to isothermal process when the temperature in the ship economy becomes the same as $T_H$ and $T_C$. We stopped the simulation when the quantity of goods on the ship is the same as initially, to make a comparison of the changes in amounts of money in the various parts of the system.}
    \label{fig:revCarnot}
\end{figure}

Figure~\ref{fig:revCarnot} {illustrates} the reverse Carnot cycle.  As {with} the forward Carnot cycle, we first demonstrate movement of money (but this time from cold to hot) with the artificial maintenance of the cold and hot temperatures to simulate the case of infinitely large mainland economies.  Secondly, we demonstrate the creation of a temperature difference from initially identical temperatures in two finite economies.

For the first demonstration, we again took $T_C=20$, $T_H=26$.  We started the ship economy at $T_S = 20$ with $G=4000$, $M=4000$.  The trader buys goods from the ship at {$3\%$} above market price, thereby increasing the money on the ship and hence its temperature.  When $T_S$ reaches $T_H$, the trader puts the ship in financial contact with the hot economy and continues to buy goods from the ship.  In this phase the resulting money flows from the ship to the hot economy.  When the {quantity} of goods on the ship reduces to $2000$, the trader disconnects the ship from the hot mainland and switches to selling goods to the ship at {$3\%$} under market price.  As a result, the {amount} {of} money on the ship falls and hence its temperature falls.  When the temperature reaches $T_C$ the trader puts the ship in financial contact with the cold economy and continues to sell goods to the ship.  The resulting money flow comes from the cold economy.  This continues until the goods on the ship returns to its initial value of $4000$, where the cycle closes. {The nett result is that money has been moved from the cold economy to the hot economy; and the trader has needed to ``spend'' (a smaller amount of) money in order to achieve this. Thus, if the trader wants, for some reason, to \emph{increase} the amount of money in the hot economy (and thus raise its temperature further), it could simply directly ``donate'' some quantity of money to the hot economy; but the trader can add money to the hot economy much more efficiently by using the reverse Carnot cycle. This can be seen as analogous to efficiency gain achieved by a heat pump; and this efficiency gain can be measured by the coefficient of performance---here the amount of money gained by the hot economy, divided by the money expended by the trader. If we focus instead on the goal of \emph{removing} money from the cold economy (and thus making it even cooler), by analogy with refrigeration, then the appropriate coefficient of performance is the amount of money lost by the cold economy, divided by the amount of money expended by the trader.}

The results of the first simulation are that in one cycle the hot economy gains $\odot 2231$ and the cold economy loses $\odot 1565$. {The difference, $\odot 2231-\odot 1565=\odot 666$ is added by the trader.
The ratio $1565/2231 \approx 0.701$.  {We can compare this with} $T_C/T_H = 20/26 \approx 0.769$, which is the theoretical maximum for this ratio, {by analogy with thermodynamics.}}
Considered as a mechanism for putting money into the hot economy, the coefficient of performance is $2231/666\approx 3.35$. Alternatively, considered as a mechanism {for removing} money from the cold economy, the coefficient of performance is $1565/666 \approx 2.35$.

For the second simulation, we started all three economies from the same temperature $T_S=T_1=T_2=20$ (we denote the mainland economies now by 1 and 2, because initially neither is hotter or colder than the other). The initial {quantity} of goods on the ship is $G=4000$ and initial money is $M=4000$.
The trader puts the ship into
financial contact with the first economy. The trader then starts buying goods from the ship at {$3\%$} above market price.  As the money on the ship increases, money flows from the ship to the first economy to keep them at equal temperatures. This money flow out of the ship economy  causes nett money flow to the first economy and hence its temperature rises.
When the {quantity} of goods in the ship economy reduces to $2000$ 
the trader disconnects the ship from the first mainland and switches to
selling goods to the ship at {$3\%$} below market price, so $T_S$ decreases and the {quantity} of goods on the ship increases.
When $T_S$ reaches the second economy's temperature (which is still $T_2=20$), the trader puts the ship into financial contact with the second economy. The trader continues to sell goods to the ship at {$3\%$} below market price so to maintain equal temperatures, money flows out from the cold economy and its temperature decreases.
When the {quantity} of goods in the ship economy reaches $4000$ 
the trader disconnects the ship from the second mainland and switches to buying goods from the ship at {$3\%$} above market price so  $T_S$ increases and the {quantity} of goods on the ship decreases.
The trader connects to the first mainland when $T_S$ reaches its temperature, and the cycle starts again. {For purposes of illustration,} we changed the threshold {quantities} of goods for the switches each time in order to spread out the resulting diagram into an expanding spiral.

Thus, by putting in a relatively small amount of money ($\odot 783 $), the trader can produce a {relatively large} temperature difference between two economies with the same initial temperature (just as in a heat pump). 
So while the trader puts in just $\odot 783$, the amount of money in the hot economy increases by $\odot 3625$ {while} that in the cold economy decreases by $\odot 2296$ {(the conservation of money requires, of course, that the money in the ship economy decreases, which it does, by $\odot 545$, {making up the difference, up to rounding error})}.  

\section{Partial derivatives}

The theory {of thermal macroeconomics} predicts various relations between partial derivatives of macroeconomic variables.  

{The simplest relations to describe are those for the variation of the ``values'' $\nu_i$ of each type of good, including money, with the {quantities} $G_j$ of goods, including money {(so considering money $M$ as the $0$-component of the vector $G$, then $\nu_0 = \beta$)}.  The theory predicts that the matrix of partial derivatives $\frac{\partial \nu_i}{\partial G_j}$  
is symmetric and negative semi-definite.  The latter is equivalent to each principal minor of the matrix having sign equal to the parity ($+$ for even, $-$ for odd) of its dimension (or being zero).  Thus the diagonal elements should be less than or equal to zero, the principal $2\times 2$ minors should be greater than or equal to zero, etc.}

To make interesting results (i.e.~{with} non-diagonal matrices) we  simulate less simple examples than CD. For realism, we keep money ``pure'',\footnote{{This is a concept introduced in \cite{CM}.  In the present context it just means that the utility functions {for each agent} depend on the amount $m$ of money {only through a multiplicative} factor of $m^{\eta-1}$ for some $\eta$.}} but we take two types of good $G_1, G_2$, with agents caring only about their sum ({so that the goods are perfect substitutes in the sense described above}),
or giving utility to one good only if matched by the other good ({the goods are perfect complements}).  {We use $u = m^{\eta-1} (g_1+g_2)^{\alpha -1}$ to capture substitutability, and $u = m^{\eta-1} \min(g_1,g_2)^{\alpha-1}$ to capture complementarity.}\footnote{To allow the extreme cases that a complements economy has no goods of one type, we have also used  $m^{\eta-1}(1+\min(g_1,g_2))^{\alpha-1}$.  To allow each good to have some utility on its own, one could multiply $u$ by a factor like $(1+|g_1-g_2|)$ or $(g_1+g_2)$, again possibly to some power.} 
  
To measure value in a general economy, as explained in Section~\ref{sec:CDmeter}, we attach a CD economy allowing movement of money and both types of good, but with artificial replacement in the target economy, and measure values there by $\nu_i = \frac{\alpha_i N}{G_i}$ in the CD economy, for $i=1,2$.

\begin{figure}[ht]
    \centering
\subfigure[\textbf{}]{
\includegraphics[width=5cm]{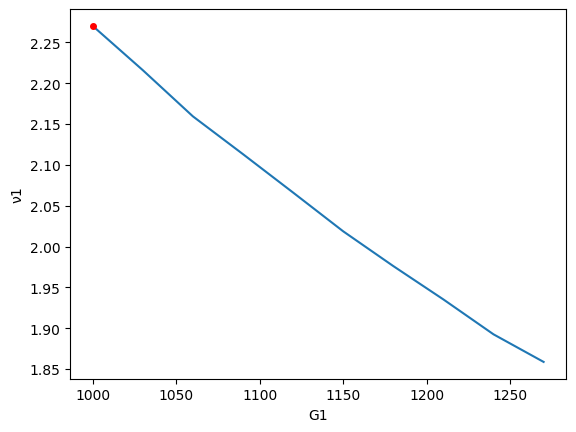}
\label{a}
}
\quad
\subfigure[\textbf{}]{
\includegraphics[width=5cm]{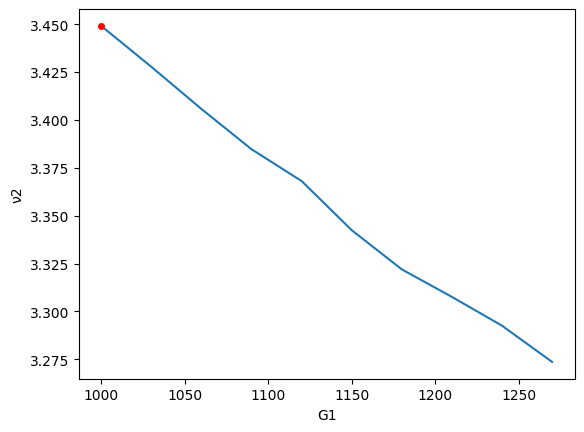}
\label{b}
}
\quad
\subfigure[\textbf{}]{
\includegraphics[width=5cm]{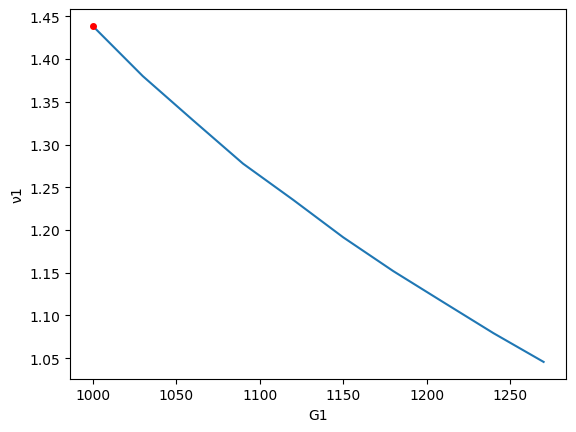}
\label{c}
}
\quad
\subfigure[\textbf{}]{
\includegraphics[width=5cm]{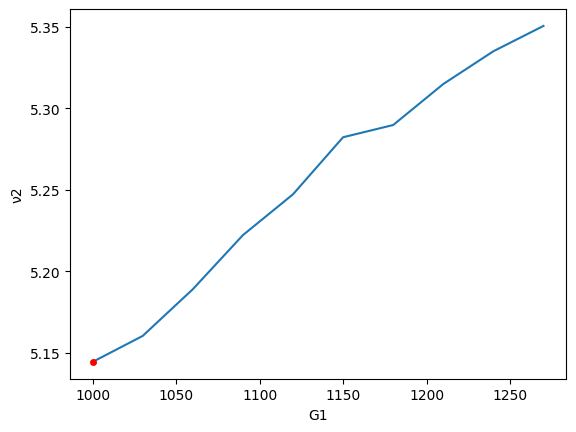}
\label{d}
}
\caption{(top row)  $\nu_1$ and $\nu_2$ as functions of $G_1$ for a substitutes economy with
$\eta=3$, $\alpha=3$, $N=1000$, $G_1 $ from 1000 to 1270,
$G_2=500$, $M =1000$;
(bottom row) $\nu_1$ and $\nu_2$ against $G_1$ for a complements economy with $\eta=3$, $\alpha=3$, $N=1000$, $G_1$ from 1000 to 1270, 
$G_2=500$, $M =1000$.
The red dots indicate where the derivative was estimated to contribute to the matrices in Table~\ref{tab:valuematrices}.}
    \label{fig:derivatives}
\end{figure}

Some results are shown in Figure~\ref{fig:derivatives}, for an economy in which goods of types 1 and 2 are substitutes (top row) and for one in which they are complements (bottom row). We see that indeed the slopes of $\nu_1$ against $G_1$ are negative in both cases, consistent with the theory.  By symmetry between types 1 and 2 in these examples, the same applies to the slopes of $\nu_2$ against $G_2$.  For the substitutes economy, the slope of $\nu_2$ against $G_1$ is also negative, but for the complements economy it is positive.  

The signs of the slopes are consistent with the interpretations. {Some comments are appropriate, however, about the values of the goods.  One might have expected that the values of two types of substitutable good should be identical for given quantities (indeed to depend only on the total), but {inspecting }the scales for $\nu_1$ and $\nu_2$ in the top row show that they are not:~{the value of the less common good, $G_2$ is substantially higher than the more common good, $G_1$}. Indeed, one can compute the relation between values and quantities of goods for this example (see Appendix~\ref{app:subs}) and the result is non-trivial.  Thus, although each agent cares about only the sum of its quantities of the two types of good, the economy cares about the quantities of the two types separately.  This merits further reflection.  Perhaps {a more realistic analysis of} substitutable goods involves other significant features {that are not currently} included in our model.  Another point for reflection is {a possible parallel with} Gibbs' paradox in statistical mechanics, where different theoretical results are obtained if one considers two types of molecule which do not differ in their physics {at the micro-level} but are distinguishable in some other way (e.g.~isotopic composition).}

In Table~\ref{tab:valuematrices}, we give numerically estimated values for the matrices of derivatives $\frac{\partial\nu_i}{\partial G_j}$, for the two types of economy at $G_1=1000$, $G_2=500$. 
{To obtain $\frac{\partial \nu_i}{\partial G_1}$ at a $[M,G_1,G_2]$ point{, we} measured $\nu_i$ at $[M,G_1,G_2]$ and at a state with $3\%$ change in $G_1$, and took the finite difference quotient.  The other partial derivatives were estimated in a similar way.

\vskip 2ex
\begin{table}[!h]
    \centering
    \begin{tabular}{cc}    
        $\begin{bmatrix} -1.793 & -0.737 \\ -0.717 & -5.274 \end{bmatrix}$ &
        $\begin{bmatrix} -1.910 & 1.099 \\ 1.116 & -12.133 \end{bmatrix}$ \\      
    \end{tabular}
    \caption{$\frac{\partial \nu_i}{\partial G_j}$ matrices ($i,j = 1,2$) for 
    (left):~a substitutes economy, and 
    (right):~a complements economy. 
    Both have $\eta=3$, $\alpha=3$, $N=1000$, $G_1 =1000$, $G_2=500$, $M =1000$. The derivatives were computed with $M$ constant.}
    \label{tab:valuematrices}
\end{table}

The value-vector $\nu$ was measured by attaching a CD economy. We took the mean $\nu$ of {the} 10,000 last samples from {the} attached CD economy as the measured value.}  
To obtain an accurate measurement of $\nu$ in a relatively short time, we let the CD economy update a small percent of the real change to reduce the fluctuations. As we also wanted to reduce the time it takes to reach equilibrium, we used a decaying step size.  At first, the update size is 1 (update $100\%$ of the change for 50000 iterations), then the step size is reduced to 0.4  of the previous value after 40000 iterations. After reducing four times, we then recorded 100,000 values of $\nu$ and took the mean as the measured $\nu$. In each cycle, the target economy trades 310,000 times with the test economy. We tested 100 cycles and took the mean to obtain the final $\nu$ in this section.

We see that the matrices are approximately symmetric (based on the measurement procedure, we could expect errors of the order of a few percent). The diagonal entries are negative and the determinants (about 8.59 and 21.85, respectively) are positive, {as predicted by the theory}.

We also checked that for these two systems the derivatives of $\nu_0=\beta$ with respect to $G_1, G_2$, and of $\nu_1, \nu_2$ with respect to $M$, are negligible:~the measured $\frac{\partial \beta}{\partial G_1}$ is -0.047 and -0.087, respectively, and $\frac{\partial \beta}{\partial G_2}$ is 0.107, -0.107, which are less than $2\%$ of the $\frac{\partial \nu_i}{\partial G_i}$. This is consistent with the theoretical result of zero for these partial derivatives for these utility functions (which are products of functions of amounts of money and quantities of goods, so money and goods are independent). 

A more sophisticated prediction of thermal macroeconomics is that, with more than one type of good, the matrix of ``flexibilities'' $\calM_{ij} = \frac{\eth \mu_i}{\eth G_j}$ is symmetric and negative semi-definite\footnote{The flexibility matrix is the inverse of the more common elasticity matrix in economics.}.  Here, the derivatives are ``compensated'' derivatives, i.e.~{any change in the {quantity} of goods is compensated by an opposite change in the amount of money, representing that the good is bought or sold at market prices. The overall change to the economy is therefore reversible and hence corresponds to no overall change in entropy}, {so can be referred to as isentropic}. 

Computation of the flexibilities for substitutes and complements economies by numerical differentiation, analogous to those for Table~\ref{tab:valuematrices}, gives the results of Table~\ref{tab:flexmatrices}.
As predicted, we see negative-definiteness and approximate symmetry again.
\vskip2ex
\begin{table}[h]
    \centering
    \begin{tabular}{cc}      
        $\begin{bmatrix} -1.197 & -1.124 \\ -1.237 & -3.015 \end{bmatrix}$ &
        $\begin{bmatrix} -0.855 & -0.449 \\ -0.449 & -6.795 \end{bmatrix}$ \\      
    \end{tabular}
    \caption{$\frac{\eth \mu_i}{\eth G_j}$ matrices for 
    (left):~a substitutes economy, and 
    (right):~a complements economy. 
    Both have $\eta=3$, $\alpha=3$, $N=1000$, $G_1 =1000$, $G_2=500$, $M =1000$. The derivatives are compensated derivatives.}
    \label{tab:flexmatrices}
\end{table}

\section{Onsager matrix}

On putting in contact two economies with a vector $\Delta$ of small differences in values of goods and money, we expect a vector $J$ of nett flows of goods and money roughly proportional to $\Delta$:
\begin{equation}
J = L \Delta,
\end{equation}
for some matrix $L$.
By analogy with physics, we call $L$ the {\em Onsager matrix}. 
The matrix $L$ will depend, of course, on the way in which the two economies are connected.  For example, if only a few agents in each are in contact with each other then $L$ will be small, whereas if all agents in each are in contact with all in the other, then it may be large.

According to thermal macroeconomic theory, the flows will reduce the value-differences (though not necessarily monotonically) and bring the two economies into equilibrium with equal values for each type of good and money.  

How could $L$ be measured? A direct but practically difficult approach would be to attempt to maintain a given value-difference vector by adjusting the flows of goods and money into one economy and out of the other (allowing positive and negative flows for different types). But it is much simpler to work in the opposite direction, from flows to value-differences. Specifically, we input equal and opposite flows $J$ to the two economies and see what time-averaged vector $\Delta$ of value-differences results. 
Assuming $L$ is invertible, this would be \begin{equation}
\Delta = K J
\end{equation}
with $K = L^{-1}$, so giving a way to measure $K$ (which could then be inverted if desired to compute $L$).
One can measure values in each economy by attaching a CD meter, as proposed in Section~\ref{sec:CDmeter}.  

According to the theory of \cite{CM}, $L$ must be positive semi-definite (psd), because entropy is produced at rate $\dot{S} = \Delta^T J = \Delta^T L \Delta$ {(where superscript $^T$ denotes transpose)}, which can not be negative.  Equivalently, $K$ must be psd because $\dot{S} = J^T K^T J$.
To make interesting examples, we test the positive semi-definiteness of $K$ on pairs of economies with substitution or complements effects (for two CD economies with any number of types of good, $K$ would be diagonal, which is less interesting).

We simulate this process by connecting either two substitutes economies or two complements economies.  We chose to connect each agent in one to each agent in the other. 
We attach CD meters to both, to measure the values of money and goods in them.

In more detail, we attach two substitutes or complements economies $A$ and $B$. {In each case, we take $N=1000, M=1000, G_1=1000, G_2= 500, \alpha=\eta=3$.  Then the value $\nu_0$ of money in each is $\beta = N\eta/M= 3$.
The values of the two types of good are measured as in the previous section and found to be $\nu_1 = 2.27, \nu_2 = 3.45$ for each substitutes economy, and $\nu_1 = 1.44, \nu_2 = 5.13$ for each complements economy (the red dots in Figure~\ref{fig:derivatives}).}
We set the flow of $[M,G_1,G_2]$ into $A$ and out of $B$ to $$J=[j,0,0], [0,j,0], [0,0,j]$$ in turn, {for some small value of $j$}. 
We used $j=0.03$ for the substitutes economies and $j=0.02$ for the complements economies,
in units of $10^{-6}$ in time {(representing one encounter)}. 
{The flux is implemented by adding or removing the given quantity to a random agent at each time-step.} 
The values of goods and money were measured by attaching CD economies to economies $A$ and $B$ after letting the system settle to equilibrium again (with the updates being applied to only the CD economies).  The resulting value-gaps were measured; for example for the pair of substitutes economies with flux $[0.03,0,0]$ the difference in value of money in $A$ and $B$ was measured to be $0.042$ and the differences in values of the two types of good were negligible.  
{The value-gaps between the substitutes economies were at most $3\%$ of the initial values, thereby making it plausible that we were in the linear regime.  For the complements economies, the value gaps were up to $8\%$ of the initial values, but it was hard to make accurate measurements for smaller fluxes, and a linearity test to be described shortly suggests that we were in the linear regime anyway.}
Then we combined the results for the different choices of $J$ to make the matrix $K$, shown in Table~\ref{tab:Onsager}.

\begin{table}[h]
    \centering
    \begin{tabular}{|c|c|}
        \hline
        \textbf{Substitutes Economy} & \textbf{Complements Economy} \\
        \hline
        $\begin{bmatrix}
        1.458 & \begin{bmatrix} 0.08 &  -0.013 \end{bmatrix} \\
        \begin{bmatrix} -0.067 \\ -0.017 \end{bmatrix} & \begin{bmatrix} 2.711 & 0.364 \\ 0.355 & 0.909 \end{bmatrix}
        \end{bmatrix}$ &
        $\begin{bmatrix}
        1.418 & \begin{bmatrix} -0.012 & -0.049 \end{bmatrix} \\
        \begin{bmatrix} -0.021 \\ -0.010 \end{bmatrix} & \begin{bmatrix} 6.117 & -0.465 \\ -0.468 & 0.979 \end{bmatrix}
        \end{bmatrix}$ \\
        \hline
    \end{tabular}
    \caption{$K$ matrices for substitutes and complements economy pairs, {using $10^{-6}$ as unit of time}. All economies have $N=1000$, $\alpha = \eta=3$, and initial $G_1=1000$, $G_2=500$, $M=1000$.}
    \label{tab:Onsager}
\end{table}
We see that $K$ is indeed psd, {and is} in fact positive-definite.  We also see the expected block structure (up to measurement error), coming from taking money independent of goods.  The signs of the off-diagonal elements are also consistent with the interpretations as substitutes and complements.

{To test linearity, we computed two other cases, $J=[0,0.03,0.03]$ and $[0,0.03,-0.03]$ for substitutes economies, and $J=[0,0.03,0.03]$, $[0,0.02,-0.02]$ for complements economies. The measured value gaps $\Delta$ for the pair of substitute economies are [0.002, 0.090,  0.037] and [-0.001, 0.067, -0.017]  for these two cases of $J$ and the theoretical results calculated from the matrix $K$  are [0.002, 0.092, 0.038] and  [0.003, 0.070, -0.017], respectively.  The  measured value gaps for a pair of complements economies are [-0.002, 0.113, 0.008] and $[-0.001, 0.134, -0.032]$ in the two cases of $J$, and the theoretical value gaps using the matrix $K$ are  [-0.001, 0.113, 0.010], [0.001, 0.132, -0.029].} 
The agreement is {pretty good}. {For larger fluxes, the value-gaps become comparable to the initial values, but surprisingly we see only slight deviations from linearity. For example, for complements economies with $J = [0,0.2,0]$, then the measured $\Delta = [-0.002,1.272, -0.108]$, whereas that given by the matrix $K$ is $[-0.002, 1.223, -0.094]$.}

A further point is that according to Onsager, if the microscopic dynamics is time-reversible then in addition $L$ (and so $K$) is symmetric.  Our examples are time-reversible so this can be tested.  {Indeed, in Table~\ref{tab:Onsager}, $K$ looks close to symmetric.} 
It would be interesting to look for failure of symmetry in some non-reversible examples too.

Many possible extensions remain for future work. For example, it would be interesting to compute the Onsager matrix for the connection of a substitutes economy with a complements economy. It will also be interesting to consider more realistic economies with many goods, which can have more complex patterns of complementarity and substitutability between goods.

\section{Behavioural models}
\label{sec:beh}

An important aspect of the theory of \cite{CM} is that it requires only rather weak axioms about the macroscopic behaviour of economic systems, {and therefore} allows much more realistic microdynamics than the standard rational utility-maximisers.  In particular, we can incorporate insights from behavioural science.  For example, agents may prefer to 
copy what their neighbours are doing.  As long as the macroscopic axioms are satisfied, the theory of \cite{CM} {applies}.
{Note, though, that these axioms will not always be satisfied. For example, in some behavioural models} thresholds can be reached, beyond which the axioms fail to be satisfied.  {In particular,} if herding effects are too strong, the assumption of unique equilibrium may fail \cite{GBB}.  {So some restrictions are required to expect a fit with our thermal macroeconomic theory.}

\begin{figure}[h]
\centering
\includegraphics[height=2in]{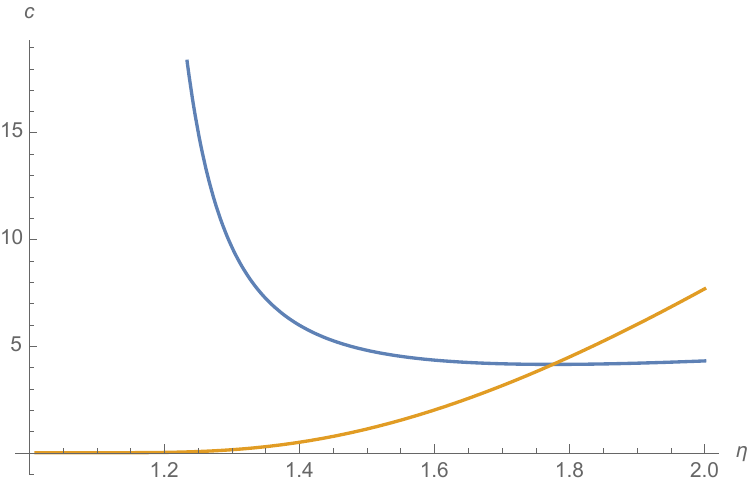}
\caption{The allowed region in the space of $(\eta,c)$, where the axioms of thermal macroeconomics hold, is that to the left of both curves ($S$ ceases to be concave to the right of the orange curve; $\beta$ is negative to the right of the blue curve).}
\label{fig:allowedregion}
\end{figure}

Here we describe some simulations with behavioural effects. 
The example we have simulated is a variant of the model in \cite{GBB}, but with the herding parameter small enough to remain in the unique phase regime.  We treat the case with only money. We call it a {\em Bouchaud economy}, because it was motivated by correspondence with J-P Bouchaud.
Each agent has utility $u(m_i) = m_i^{c \bar{m}^2}$, where $\bar{m}$ is the average money per agent in the economy and $c>0$ is a parameter.
When isolated, this economy is identical to a CD economy with $\eta = 1+ c M^2/N^2$, but when connected to a trader or another economy, we see somewhat different behaviour.  In particular, its temperature is not the CD result $\frac{M}{N\eta}$, because changing the amount of money in the economy also changes $\eta = 1 + c\bar{m}^2$.  A formula for the temperature (or actually its reciprocal $\beta$) is derived in Appendix~\ref{app:Becon}):
\begin{equation}
\beta = \frac{1}{\bar{m}} + c\bar{m} [1+2\log\bar{m} + 2\psi(\eta)-2\log\eta],
\label{eq:betaB}
\end{equation}
where $\psi$ is the digamma function $\Gamma'/\Gamma$.
To satisfy the axioms of \cite{CM}, one must restrict the parameters $c,\bar{m}$ (or equivalently, $c,\eta$) to the region shown in Figure~\ref{fig:allowedregion}.

\begin{figure}[h]
\centering
\includegraphics[height=1.6in]{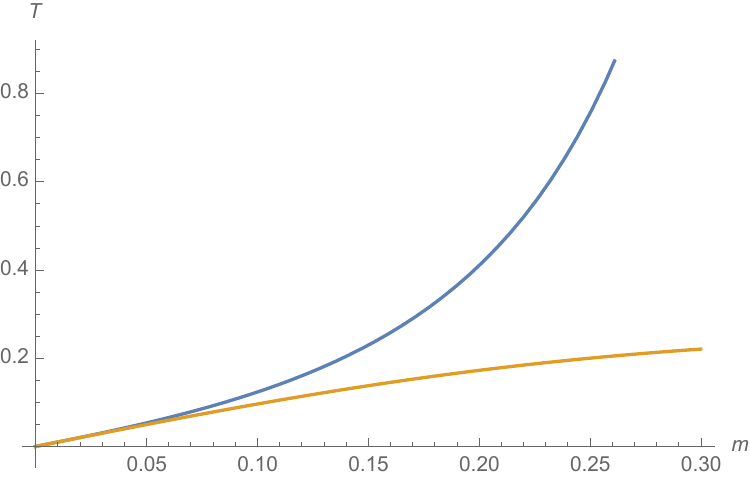}
\caption{Temperature $T$ against average money $m (= \bar{m})$ for the Bouchaud economy with $c=4$ in blue, compared to the approximation by a CD economy with $\eta = 1 + c\bar{m}^2$ in orange.}
\label{fig:temp}
\end{figure}

To obtain an idea of the functional form, one can use the approximation
$$\psi(\eta)-\log\eta \sim -\gamma + (\frac{\pi^2}{6}-1)(\eta-1)$$
as $\eta \to 1$, where $\gamma \approx 0.577$ is the Euler-Mascheroni constant.  
So
\begin{equation}
\beta \sim \frac{1}{\bar{m}} +c\bar{m}[1+2\log\bar{m} - 2\gamma + 2(\frac{\pi^2}{6}-1)c\bar{m}^2],
\label{eq:betaBapprox}
\end{equation}
as $\bar{m} \to 0$.
Thus $\beta \sim \frac{1}{\bar{m}}$ for $\bar{m}$ small (as for a CD economy with $\eta=1$), but lies below this for $\bar{m}$ small and positive.
Equivalently, $T = \frac{1}{\beta} \sim \bar{m}$ for $\bar{m}$ small and increases above this as $\bar{m}$ increases (whereas the CD approximation  $\frac{\bar{m}}{\eta}$ decreases below $\bar{m}$). 
Figure~\ref{fig:temp} shows the comparison between the exact formula (\ref{eq:betaB}) and the CD approximation for the case $c=4$.

\begin{figure}[h!]
\centering
\includegraphics[height=1.5in]{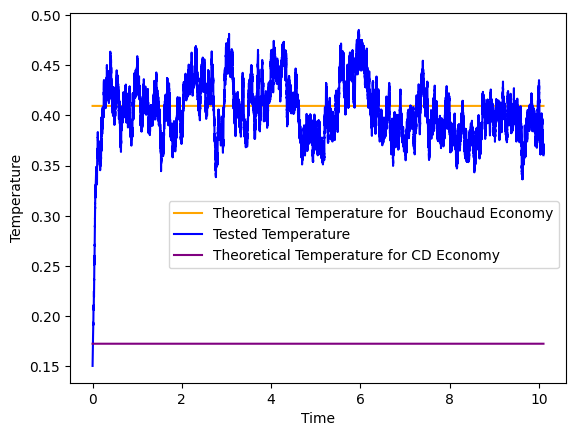}
\includegraphics[height=1.5in]{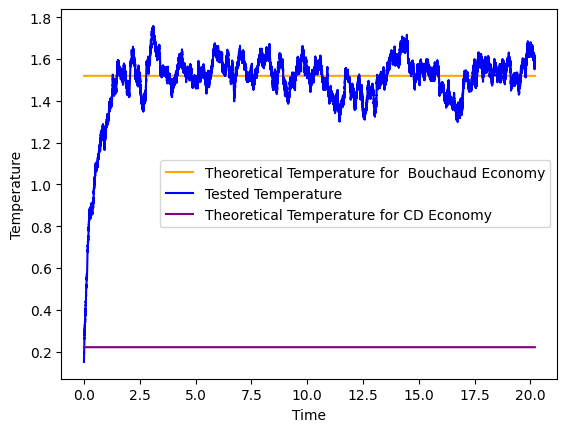}
\includegraphics[height=1.5in]{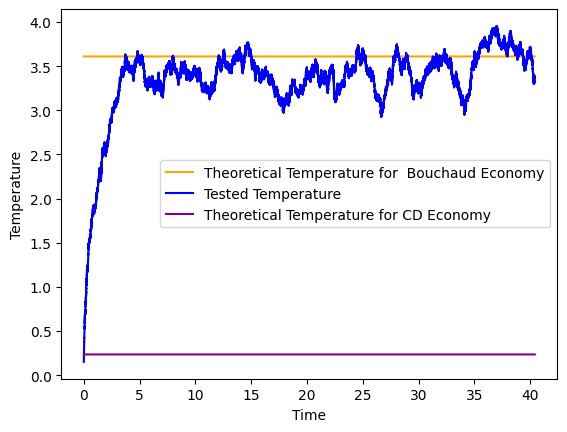}
\includegraphics[height=1.5in]{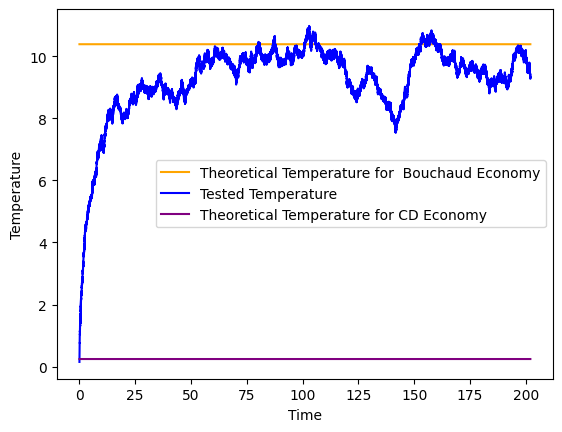}
\caption{Temperature of a Bouchaud economy with $c=4$, $N=100$, and $M=20,30,35,40$,  measured by a CD economy as a function of time.}
\label{fig:temp2}
\end{figure}

We confirm the formula (\ref{eq:betaB}) for the temperature of a Bouchaud economy by measurement by a CD meter. 
Examples are shown in  Figure~\ref{fig:temp2}. 
They fit well except for the case $M=40$, which is perhaps too close to the boundary of the allowed region in parameter space.

\begin{figure}[h!]
\centering
\includegraphics[height=1.9in]{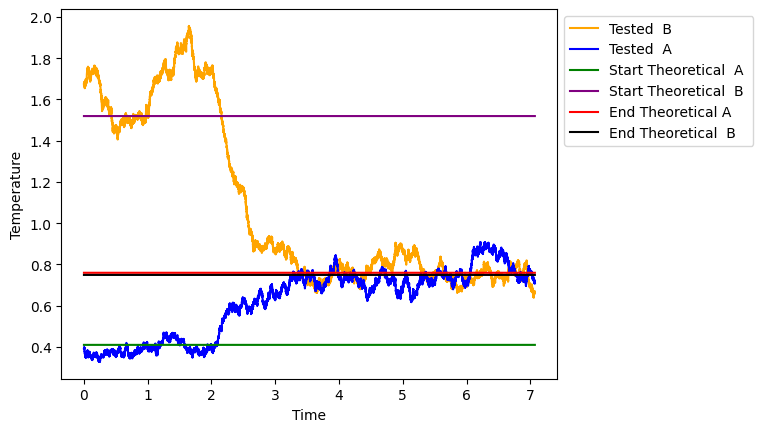}
\caption{Temperatures of two Bouchaud economies put into financial contact at time
$2$, as functions of time. The parameters are $c=4, N = 100$ for each, and initial $M_A = 20, M_B = 30$.}
\label{fig:temp3}
\end{figure}

Finally, we put two Bouchaud economies in financial contact and confirm that nett money flows from hotter to cooler until the temperatures equilibrate.  This is shown in Figure~{\ref{fig:temp3}}.

Variants of this economy could be studied.  For example, the agents might be divided into cells and the average $\bar{m}$ to be used in their utility function {calculated} only over the agent's cell; {or} there could be a concept of distance between agents and the average $\bar{m}$ would be taken with respect to a decaying function $f$ of distance $d$ from the given agent.  A special case of the latter would be to take $f(d)$ to be zero for distance $d$ exceeding some threshold, which would speed up its computation.  A downside is that we don't have a theoretical formula for the entropy or temperature to compare with in these cases.

\section{Conclusion}

To summarise, we have confirmed {a variety of} predictions of thermal macroeconomic theory on some simple simulated micro-economies in a range of contexts.  Notably, after checking that our economies go to a unique equilibrium subject to given amounts of conserved quantities, we have demonstrated that in financial contact, money flows from hotter to cooler; in trading contact, goods flow from lower to higher price; on putting two economies in contact, the total entropy does not decrease, though the entropy of one {economy} may decrease; a trader can make money out of a temperature difference between two economies; a trader can move money against a temperature difference by putting in a relatively small amount; the matrix of derivatives of values with respect to {quantities} of goods is negative semi-definite (nsd) and symmetric; the matrix of compensated derivatives of prices with respect to {quantities} of goods is also nsd and symmetric; the Onsager matrix giving flows for small value-differences between two economies is positive semi-definite.
We tested cases of economies with substitution effects or complements effects and also economies with social behavioural effects.

{There are many possible} directions for future work. {For example,} for our models of encounters, we chose the encounter-rate matrix $k$ to be symmetric, but one could use non-symmetric $k$ to allow for a difference in outcome for agents $i,j$ depending on which was the first mover.  Agents could have different utility functions according as they are first mover or not. This would in general break reversibility of the Markov process, making explicit formulae for the stationary distribution inaccessible, but could still be a valid model. {It would be valuable to simulate multiple goods and currencies, with potential restrictions on flows and exchanges}. In the context of partial derivatives, it would also be interesting to attempt to confirm our version of Le Chatelier's principle (see \cite{CM}), {which has been viewed as of central importance by the founder of modern mathematical economics, Paul Samuelson \cite{S72}. Moreover, the range of behavioural factors that could be incorporated into the models is large, including, for example, utilities depending on changes as well as absolute values; more ``heuristic'' rules for agents agreeing to an exchange; and more restrictive behaviourally-inspired rules capturing which pairs of agents can, or choose to, make exchanges within an economy or between economies. In the longer term, we hope to extend the thermal macroeconomic theory described in \cite{CM} beyond exchange economies to include multiple currencies, production, consumption, credit and so on. Such extensions could naturally be tested with simulated microeconomies building on the work outlined here.}

\section*{Acknowledgements}
This paper is an extension of preliminary computational simulations carried out at the University of Warwick by Daniel Sprague in 2011 and Shafi Sardar in 2022/3.
{We are grateful for comments from Jean-Philippe Bouchaud and grants from Warwick Business School Primary Data Fund and the University of Warwick Research Development Fund for financial support October 2023 to September 2024. For the purpose of open access, the authors have applied a Creative Commons Attribution (CC-BY) licence to any Author Accepted Manuscript version arising from this submission.}

\section*{Data Availability}
The code for the simulations reported here is available on github at \\ \texttt{https://github.com/YihangLuo99/Thermoeconomics}.

\section*{Appendices}
\appendix

\section{Selection of encounters}
\label{app:A}
If the encounter matrix $k$ has all $k_{ij}=1$ (except on the diagonal) then one simply selects a pair $(i,j)$ uniformly at random.

For a more general encounter matrix $k$, a simple way to select the next encounter is to label the agents in some order (if labelled by integers then use the natural order) and then order the pairs $(i,j)$ by $(i',j') \le (i,j)$ if $i'\le i$ and for $i'=i$ then $j' \le j$.  Let $K = \sum_{i,j} k_{ij}$ and $A_{ij} = K^{-1} \sum_{(i',j')<(i,j)} k_{i',j'}$, which form a nowhere decreasing sequence.  Select a random number $r$ uniformly in $[0,1]$.  Let $(i,j)$ be the greatest (in the order) such that $A_{ij} < r$.  Then choose encounter $(i,j)$.

If it is desired to keep track of real time, then one can choose an exponentially distributed time interval with mean $K^{-1}$ (see Appendix~\ref{app:B}).  But in practice we will just count the number of encounters and divide by $K$.

\section{Sampling from a probability distribution}
\label{app:B}
For easy probability distributions, one can sample directly, using a random number generator that is assumed to produce independent samples uniformly in $[0,1]$.  We say these are from $U(0,1)$.

The easiest case is of course a uniform distribution, say on an interval $[A,B]$.  Then given sample $r$ from $U(0,1)$, let $x = (1-r)A + rB$.  It is a sample from $U(A,B)$.

More generally, for a probability distribution $P$ on $\R$ with density $\rho$, let $C(x) = \int_{x'<x} \rho(x') dx'$, called the cumulative distribution function.  $C$ is nowhere decreasing and goes from 0 to 1.  
Let $C^{-1}$ be the inverse function to $C$ (more precisely, in case $C$ has intervals of constancy, let $C^{-1}(r)$ be the supremum of $x$ such that $C(x)<r$) and suppose it is computable:~we say $P$ is an {\em easy} distribution. Select a random number $r$ uniformly in $[0,1]$ and let $x = C^{-1}(r)$.  This is a random sample from $P$.

An example of easy probability distribution is the standard exponential one with density $\rho(x) = e^{-x}$ for $x\ge 0$.  Then $C(x) = 1-e^{-x}$, so $C^{-1}(r) = -\log(1-r)$.
Thus, $-\log(1-r)$ is a random sample from the exponential distribution.  Since $1-r$ is uniformly distributed on $[0,1]$ for $r$ uniformly distributed on $[0,1]$, one can equally well select $-\log r$ as long as one is consistent.

Another example from which it is easy to sample is a standard Gamma distribution Gamma($k$) with integer parameter $k$:~$\rho(x) = \Gamma(k)^{-1} x^{k-1}e^{-x}$ on $x\ge 0$.  The case $k=1$ is just the exponential distribution already treated. The cumulative distribution for general $k$ does not have a nice inverse, but for integer $k$ one can sample by choosing $k$ independent random numbers $r_1,\ldots r_k$ uniformly in $[0,1]$ and letting $x = -\sum_{j=1}^k \log r_j$.  For Gamma($k,m$) with mean $m$, i.e.~$\rho(x) \propto x^{k-1}e^{-kx/m}$ (conventions differ about notation for this), let $x = -m\sum_{j=1}^k \log r_j$. 

Another easy example is the standard Beta($a,b$) distribution, for which the density is $\rho(x) = Z^{-1} x^{a-1} (1-x)^{b-1}$ for $x \in [0,1]$ with $a,b$ positive integers ($Z = B(a,b)$ is the associated normalisation constant).   Let $A$ be a sample from Gamma($a$) and $B$ be an independent sample from Gamma($b$). Then $\frac{A}{A+B}$ is a sample from Beta($a,b$).  It is easy to sample from scaled versions, e.g.~for $\rho(x) \propto x^{a-1}(X-x)^{b-1}$ on $[0,X]$, take $x=\frac{XA}{A+B}$.

Products of easy distributions are easily sampled from by taking samples of each component independently.

To sample from a general probability distribution $P$ on $\R^n$ with density $\rho$, a common method is an iteration known as the Metropolis-Hastings (MH) algorithm (though it appears credit should really also be given to the Rosenbluths and Tellers).
It depends on choosing a ``proposal'' distribution $g(x'|x)$ for the next point $x' \in \R^n$ given the current one $x$, and accepting $x'$ with an ``acceptance'' probability $A(x',x)$, else rejecting it and staying at $x$.  The proposal distribution must be easy to sample from, in the above sense.  Thus for example it can be a product of independent easy distributions in the coordinate directions.

A suitable choice of the acceptance probability is given by
$$A(x',x) = \min\left(1,\frac{\rho(x')}{\rho(x)} \frac{g(x|x')}{g(x'|x)}\right).$$
This ensures that the MH iteration has $\rho$ as a stationary density.  If $g$ is chosen so that the MH iteration explores the whole available space (one can make this mathematically precise) then $\rho$ is the only stationary density and with probability 1 the empirical distribution of the iterates from an arbitrary initial condition converges to $\rho$ (in the dual to the space of continuous functions).
Thus one can approximately sample from $\rho$ by iterating MH long enough and taking the last sample (but the stopping time must be chosen independent of the position).

There are practical questions about what counts as long enough.  It is good to take $g(x'|x)$ to approximate $\rho(x')$, to avoid the acceptance probability being small too often.  A common choice of $g$ is to take $x'$ uniform over a rectangle of given dimensions centred on $x$, but that is not particularly efficient (especially if the rectangle is small, because then it takes a lot of steps for the MH algorithm to explore the space).  In our context, we will often take Beta distributions for $g$ (independent of the current point).

There are methods that produce a perfect sample from the stationary distribution of a mixing Markov chain by a clever stopping criterion, e.g.~\cite{H}, but we will not be so sophisticated.

An interesting feature of the MH algorithm is that one doesn't have to normalise $\rho$:~the only place the probability density enters is in the acceptance probability, which depends only on the ratio of $\rho$ at the two points. This is an advantage when the normalisation is not easily computable.

\section{Substitutes economy}
\label{app:subs}

Here we derive the relation between the quantities of goods in a substitutes economy and the values of the goods.

By reversibility of the dynamics with respect to the measure with density $\prod_i (g_i+h_i)^{\alpha-1}$ on the bisimplex $\sum_i g_i = G, \sum_i h_i = H$, $g_i, h_i \ge 0$, the normalisation of this measure is a stationary probability distribution for the exchange process, and for a connected exchange network it attracts all initial probability distributions.

By a standard argument in statistical mechanics, the relation between quantities of goods and their values in the thermodynamic limit can be computed from the canonical partition function $$Z_c(\nu_g,\nu_h) = \int \prod_i (g_i+h_i)^{\alpha-1} e^{-(\nu_g g_i + \nu_h h_i)} \prod_i dg_i dh_i$$
over all $g_i, h_i \ge 0$ by
$$G = -\frac{\partial}{\partial \nu_g} \log Z_c, \ H = -\frac{\partial}{\partial \nu_h} \log Z_c.$$
The integral $Z_c$ is a product of 2D integrals, namely the $N^{th}$ power of
$$\int (g+h)^{\alpha-1} e^{-(\nu_g g + \nu_h h)} dg\, dh.$$
Change variables to 
$$\sigma = g+h,\ \delta = (g-h)/2$$
to obtain
$$\int \sigma^{\alpha-1} e^{-(\nu_\sigma \sigma + \nu_\delta \delta)} d\sigma d\delta$$ 
over $\sigma\ge 0$, $\delta \in [-\sigma/2,+\sigma/2]$,
where
$\nu_\sigma = (\nu_g + \nu_h)/2, \nu_\delta = \nu_g-\nu_h$.
Integrating with respect to $\delta$ produces
$$\int_0^\infty \sigma^{\alpha-1} e^{-\nu_\sigma \sigma}\, \frac{e^{\nu_\delta \sigma/2} - e^{-\nu_\delta \sigma/2}}{\nu_\delta} \, d\sigma.$$
This integrates to
$$\Gamma(\alpha) \frac{\nu_g^{-\alpha}-\nu_h^{-\alpha}}{\nu_h-\nu_g}.$$
So
$$\frac{G}{N} = \frac{\alpha \nu_g^{-\alpha-1}}{\nu_g^{-\alpha}-\nu_h^{-\alpha}} + \frac{1}{\nu_g-\nu_h},\ \frac{H}{N} =  \frac{\alpha \nu_h^{-\alpha-1}}{\nu_h^{-\alpha}-\nu_g^{-\alpha}} + \frac{1}{\nu_h-\nu_g}.$$

This has the following interesting limiting regimes.  If $\nu_h \gg \nu_g$ then $G \sim N\alpha/\nu_g$, $H \sim N/\nu_h$, so there is relatively little of $H$, and $G$ behaves like a CD good with exponent $\alpha$.  Inverting this, if $G\gg H$ then $\nu_g \sim N\alpha/G$ and $\nu_h \sim N/H$, so we see also that $H$ behaves like a CD good but with exponent 1 instead of $\alpha$.

If instead, $\nu_g=\nu_h$, then by careful treatment of the limit $\nu_g \to \nu_h$, $G=H = N(\alpha+1)/2\nu_g$, thus $G+H$ behaves like a CD good with exponent $\alpha+1$.  We can turn this round to obtain that when $G=H$ then $\nu_g=\nu_h = \frac{N(\alpha+1)}{G+H}$.

A similar analysis can be carried out for our complements economy, to be reported elsewhere.

\section{Bouchaud economy}
\label{app:Becon}

Here we derive formulae for the entropy and temperature of a Bouchaud economy and the region of parameter space {for which thermal macroeconomic theory is applicable.}

We consider economies in which the utilities of agents for money $m$ are $u(m) = m^{\eta-1}$ with $\eta$ a function of the mean money $\bar{m}$ per agent.  They can be called ``mean-field'' economies.  We call {\em Bouchaud economy} the case $\eta = 1+c\bar{m}^2$ for some constant $c>0$.

For a closed economy with $N$ agents and total money $M$, $\bar{m}$ is constant $M/N$.  Assuming the graph of encounter rates between its agents is connected, as in \cite{CM} there is a unique and globally attracting stationary probability distribution with density $\rho = \frac{1}{Z} \prod_i m_i^{\eta-1}$ with respect to $\delta(M-\sum_i m_i) \prod_i dm_i$ on the simplex $m_i \ge 0$, $\sum_i m_i = M$.  The normalisation constant $$Z(M) = \frac{M^{N\eta-1}\Gamma(\eta)^N}{\Gamma(N\eta)}, $$
is called the ``partition function'' in statistical mechanics.

Following the general procedure in statistical mechanics, the entropy per agent in the limit $N\to \infty$ with $\bar{m}$ constant is
$$s(\bar{m}) = \lim_{N\to \infty} \frac{1}{N} \log Z(N\bar{m}).$$
Justification for this in the context of the Lieb-Yngvason axioms used in \cite{CM} will be given elsewhere.

It follows that
$$s(\bar{m}) = \eta \log\bar{m} + \log\Gamma(\eta) -\eta \log\eta + \eta. $$
In the case $\eta$ constant ({as in} a CD economy), this can be truncated to $s(\bar{m})= \eta \log \bar{m}$, because entropy is defined only up to a constant.  But when $\eta$ depends on $\bar{m}$ the remaining terms are essential for {predicting} how the economy behaves on connection to {other economies or a trader}.

In particular, when $\eta$ depends on $\bar{m}$ the temperature $T$ deviates from the CD formula $\bar{m}/\eta$.  Explicitly, the coolness (inverse temperature)
$$\beta = \frac{\partial s}{\partial \bar{m}} = \frac{\eta}{\bar{m}} + (\psi(\eta)-\log\eta+\log\bar{m})\eta'(\bar{m}),$$
where $\psi$ is the Digamma function $\Gamma'/\Gamma$.
For the case $\eta = 1 +c\bar{m}^2$, this can be written as
$$\beta = \frac{\eta}{\bar{m}} + 2c\bar{m}(\psi(\eta)-\log\eta+\log\bar{m}),$$
which can be rearranged to give equation (\ref{eq:betaB}).

Two issues come to light from the formula.  Firstly, $\beta$ may be negative.  This happens where $$c > \frac{\eta-1}{\eta^2} \exp\left(\frac{\eta}{\eta-1} + 2\psi(\eta)\right),$$
that is to the right of the blue curve in Figure~\ref{fig:allowedregion}.
Secondly, $\frac{\partial \beta}{\partial \bar{m}}$ may be positive.  This happens if
$$c> (\eta-1) e^{-F(\eta)},$$ where
$$F(\eta) = -2\psi(\eta)+2\log\eta-4(\eta-1)\psi'(\eta) - \frac{4}{\eta}+\frac{\eta}{\eta-1}.$$
namely to the right of the red curve in Figure~\ref{fig:allowedregion}.  Such parameter values violate some of the axioms of \cite{CM}, so we exclude them from our simulations.

\end{document}